\documentclass[9pt,shortpaper,twoside,web]{ieeecolor}
\usepackage{generic}
\usepackage{cite}
\usepackage{amsmath,amssymb,amsfonts}
\usepackage{algorithmic}
\usepackage{graphicx}
\usepackage{textcomp}
\usepackage{float}
\usepackage{amsmath}
\usepackage{breqn}
\usepackage{multirow}
\def\BibTeX{{\rm B\kern-.05em{\sc i\kern-.025em b}\kern-.08em
    T\kern-.1667em\lower.7ex\hbox{E}\kern-.125emX}}
\begin{document}
\title{Piece-wise Matching Layer in Representation Learning for ECG Classification}
\author{Behzad Ghazanfari, Fatemeh Afghah, Sixian Zhang\\\textit{School of Informatics, Computing and Cyber Systems}\\\textit{Northern Arizona University,  Flagstaff, AZ 86011} \\ \thanks{This material is based upon work supported by the National Science Foundation under Grant Number 1657260. Research reported in this publication was also supported by the National Institute On Minority Health And Health Disparities of the National Institutes of Health under Award Number U54MD012388. }\thanks{This work has been submitted to the IEEE for possible publication. Copyright may be transferred without notice, after which this version may no longer be accessible.}}

\maketitle

\begin{abstract}
%\magenta{magenta: comments/questions}\\
%\red{red:remove}\\
%\blue{blue:new text}

This paper proposes piece-wise matching layer as a novel layer in representation learning methods for electrocardiogram (ECG) classification. Despite the remarkable performance of representation learning methods in the analysis of time series, there are still several challenges associated with these methods ranging from the complex structures of methods, the lack of generality of solutions, the need for expert knowledge, and large-scale training datasets. We introduce the piece-wise matching layer that works based on two levels to address some of the aforementioned challenges. At the first level, a set of morphological, statistical, and frequency features and comparative forms of them are computed based on each periodic part and its neighbors. At the second level, these features are modified by predefined transformation functions based on a receptive field scenario. Several scenarios of offline processing, incremental processing, fixed sliding receptive field, and event-based triggering receptive field can be implemented based on the choice of length and  mechanism of indicating the receptive field. We propose dynamic time wrapping as a mechanism that indicates a receptive field based on event triggering tactics. The proposed layer outperforms the common kernels and pooling on high-level representations of ECG classification. This layer offers several advantages including 1) it enables the learning from few instances for training and handling long quasi-periodic time series, 2) the layer followed by a simultaneous multiple feature tracking mechanism that does not require any annotated data or expert knowledge in quasi-periodic time series, and 3) the proposed framework can provide and handle several scenarios like online processing and dynamic data-driven application systems. To evaluate the performance of this method in time series analysis, we applied the proposed layer in two publicly available datasets of PhysioNet competitions in 2015 and 2017 where the input data is ECG signal.  These datasets present a good example of  quasi-periodic time series signals with a small number of training samples, highly imbalanced classes, long time series recordings, and non-linear and complex patterns. We compared the performance of our method against a variety of known tuned methods from expert knowledge, machine learning, deep learning methods, and the combination of them.
The proposed approach improves the state of the art in two known completions 2015 and 2017 around 4\% and 7\% correspondingly while it does not rely on in advance knowledge of the classes or the possible places of arrhythmia.

\end{abstract}


\section{Introduction}
\label{sec:introduction}

Time series analysis is one of the most known problems in data science and data mining with applications in different domains including finance, industry, communication, astronomy, and healthcare \cite{yang200610,fu2011review}. Time series classification and modeling are among the key challenges in this domain. Time series modeling methods  mostly focus on time series forecasting \cite{box1970distribution,alwan1988time,brockwell2002introduction}, while the goal of time series classification problem is to classify the data points as discussed in this paper. 

One of the key challenges of machine learning methods that are not based on deep learning is to learn a proper representation of time series in such a way to capture the order and correlation of extracted features while avoiding over-fitting. Two common representation approaches in time series analysis are the frequency-domain and time-domain feature extraction methods that can reduce the number of features. There are several supervised and unsupervised approaches alongside feature transformation and extraction that have been proposed for time series classification \cite{fayyad1996data,chan1999efficient,fu2011review,goroshin2015unsupervised}. However, these methods generally have issues to scale up well with long time series, length-variant time series inputs, or handling the noise properly \cite{langkvist2014review,zhang2018deep,fawaz2019deep}. 

\begin{table*}[t]
\caption{Description of 2015 PhysioNet computing in cardiology challenge's dataset \cite{clifford2015physionet}. }

\centering
\resizebox{2\columnwidth}{!}{
\begin{tabular}{|c|c|c|c|c|}
\hline
\textbf{Arrhythmia type} & \textbf{Arrhythmia Definition} &   \textbf{\# Patients} &  \textbf{\# False Alarm} & \# \textbf{True Alarm}\\
\hline\hline
\textbf{ASY} & No QRS for 4 seconds. & 122 & 100 & 22\\\hline
\textbf{EBR} & Heart rate lower than 40 bpm for 5 consecutive beats. & 89 & 43 & 46\\\hline
\textbf{ET} & Heart rate higher than 140 bpm for 17 consecutive beats. & 140 & 9 & 131\\\hline
\textbf{VF} & Fibrillatory, flutter, or oscillatory waveform for at least 4 seconds. & 58 & 52 & 6\\\hline
\textbf{VT} & 5 or more ventricular beats with heart rate higher than 100 bpm. & 341 & 252 & 89\\
\hline\hline
\textbf{Total} &  & 750 & 456 & 294\\
\hline
\end{tabular}
}
\label{table:2015_summary}
\end{table*}

\begin{table*}[t]
\caption{Description of 2017 PhysioNet computing in cardiology challenge's dataset \cite{clifford2017af}.}
\centering
\resizebox{2\columnwidth}{!}{
\begin{tabular}{|c|c|c|}
\hline
\textbf{Class} & \centering \textbf{Description} & \textbf{\# Samples}\\
\hline\hline
\textbf{Normal rhythm} & Normal heart beats.& 5050\\
\hline
\textbf{AFib} & 
Tachycardia (HR $>$ 100 bpm), ectopic QRS complex site, the absence of P wave, and the presence of fibrillatory wave. & 738\\
\hline
\textbf{Other rhythm} & All non-AF abnormal rhythms. & 2456\\
\hline
\textbf{Noise} & Too noisy to classify. & 284\\
\hline\hline
\textbf{Total} & & 8528\\
\hline
\end{tabular}
}
\label{table:2017_summary}
\end{table*}

Representation learning has led to outstanding performance in image, speech, and natural language processing (NLP). Convolutional neural networks (ConvNets) is   one of the most known branches in representation learning which is based on several layers that learn features from large input space through layers from low-level to higher forms \cite{lecun1998gradient}. ConvNets have been applied in many applications from image, text, and time series. Recurrent neural networks (RNNs) and its varieties mostly are applied in sequential type data such as text, audio, and  time series data since they can learn temporal dependencies among the features. 
 The success of deep learning in image processing inspired  considerable research to adopt layered feature learning techniques in other domains including time series analysis \cite{yu2010deep,langkvist2014review,zhang2016unsupervised,fawaz2019deep}. Despite the powerful performance of deep learning techniques in time series analysis, there are still some factors to limit their performance toward achieving the expected benchmarks. One of the key challenges in time series analysis compared to image processing is that the input data is not stationary. Also, non-linear patterns inside of each periodic can have long time dependencies over several periods.

In summary, the majority of current representation learning methods that are used in time series analysis were originally designed for image processing, NLP, and text mining domains that are based on elements such as ConvNets or RNNs. One of the known challenges with current deep representation methods is that they cannot be properly trained when there exists only a limited number of training instances as they include a large number of layers. Furthermore, these methods cannot easily handle imbalanced datasets or length-variant time series inputs.
These approaches cannot work well when dealing with long quasi-periodic signals with variable lengths to learn classes in which each instance is composed of several hundred similar periodic parts. Furthermore, in certain applications such as remote patient monitoring, the collected quasi-periodic time series data such as Electrocardiogram (ECG) signals experience considerable noise and artifact distortions, and the signals carry a large number of possible patterns which differentiates the quasi-periodic time series classification such as ECG classification from traditional NLP and image processing domains. 

In periodic time series like ECG signals from one side, there is a potential to segment the signals to sub-parts that are repeated through the signal; however, it is challenging to extract subtle differences among several hundred different periods of the signal. Quasi-periodic time series are commonly seen in biomedical and economic applications. ECG is a quasi-periodic time series with a complex morphology that plays an important role in diagnosis and therapy of several diseases. 

The main contribution of piece-wise matching layer is to extract a set of low-level features at the first level of each beat, periodic part, and its neighbor as a local perspective. Then, we apply several functions including mode, median, entropy, and Kullback–Leibler (KL) divergence on some of the first level features in a receptive field scenario as a broad perspective. These two level features are used instead of kernel matrix and pooling layers to measure spatial and temporal representations. In other words, we define the second level in the proposed layer based on some functions such as mode, median, entropy, and  (KL) divergence in an inline manner to measure the relative spatial and temporal features of the receptive fields. In fact, the receptive field provides wider views to capture the similarities and differences of the first level features relative to each other. Thus, we consider the receptive field as a way to implement the proposed multi-level feature learning in different scenarios including offline processing, incremental processing, fixed sliding receptive field processing, and event-triggering receptive field. We utilize dynamic time wrapping (DTW) as a mechanism that dynamically indicates the place and the length of the receptive field for the event-triggering receptive field scenario.

In ECG signal monitoring and diagnosis applications, we often deal with imbalanced datasets of long time series where a limited number of training instances are available. While the proposed method is generic and can be applied to any quasi-periodic time series signal, we applied it in two known ECG classification challenges 
in order to evaluate the performance of the piece-wise matching layer. The extracted features using the proposed piece-wise matching layer improve the accuracy of ECG classification considerably. The proposed approach can provide a robust and efficient pattern mining approach for ECG classification  which does not rely on expert knowledge.

%\magenta{repetitive} The proposed functions that are applied in the second level on a receptive field depending on a scenario are Shannon and log energy entropy, mode, median, and KL divergence. We propose four scenarios including incremental processing, offline processing, fixed sliding receptive field, and event-triggered receptive field. We 

In summary, the piece-wise layer calculates spatial and temporal features and the comparative forms of them in two levels. In the first level, some features are calculated from each beat as the periodic part and its neighbors. In the second level, some functions are applied on a sequence of some of the first level features -the receptive field. The performance of the proposed method is shown in two challenges on ECG classification with public datasets \cite{PhysioNet15,PhysioNet17} that covers some known issues in the ECG signals while a variety of different approaches from expert knowledge, classical machine learning, deep learning, and the combination of them are tuned and applied for these datasets. These datasets contain long time series with seventy thousand time steps, the length of signal recordings is variant, the datasets are highly imbalanced, and the approaches that obtain top-tier results in each challenge using completely different techniques.
In the first challenge \cite{PhysioNet15}, the proposed approach uses only one lead of the signal and does not use in advanced knowledge about the place of the arrhythmia or human rules of arrhythmia. Most of the other approaches that got top-tier results are based on three leads ECG signal and they use some level of expert knowledge about each arrhythmia. In the second challenge \cite{PhysioNet17}, the proposed approach has a basic architecture of bidirectional long short-term memory and achieves the state of the art without applying human knowledge of the AFib patterns, or knowing about which parts of the signals should be selected for training. In  conclusion,  the approach utilizes the same features and structure in both of these challenges while achieves the state of the art in both cases even in comparison to approaches that are tuned for each challenge or using expert knowledge depend on the challenge.

%\magenta{this first subsection is more like a "background section". I suggest to change the title to "Background on Representation Learning in Time Series"}\green{done}
\section{Time Series Classification}
The common trend of time series classification techniques is based on supervised and unsupervised learning. These learning approaches use frequency-domain and time-domain feature extraction methods or distance measures such as DTW. Among these works, ensemble classifiers including  \textit{Collective Of Transformation-based Ensembles} (COTE) which is based on 35 classifiers on different time series representation \cite{bagnall2015time,fawaz2019deep} or HIVE-COTE \cite{bagnall2017great} which utilizes 37 classifiers offer better performance. These traditional methods do not scale up well on long quasi-periodic time series such as the physiological signals with ten thousand time steps or do not offer the expected performance on the highly noisy signals. Moreover, the ensemble classifiers need a large set of training instances and a considerable processing time.

\subsection{Background on Representation Learning in Time Series}
RNN and its variants as a class of representation learning approaches have the potential to capture the temporal relations in the sequence data and time series data. In RNNs, the network output at each time step is impacted by the shared weights of the previous time steps, where a back-propagation through time (BPTT) approach is utilized to update these weights. Since RNNs suffered from the vanishing gradient problem \cite{pascanu2012understanding}, long short-term memory (LSTM) \cite{hochreiter1997long} were proposed to provide learning of longer time dependencies.  LSTM takes advantage of three control gates, input, forget, and output, to decide about the information for memory cells that make handle longer time dependencies.  However, LSTM still cannot scale up to the signals with ten thousand time steps \cite{fawaz2019deep}.

ConvNets as a successful class of representation learning techniques can automatically identify the important features in high dimensional datasets and have been mostly utilized in image processing applications. ConvNets can capture the local relations in the images by  feature extraction based on local partitioning of the images through multiple layers, where the first layers' filters are responsible for extracting primitive elements such as the edges, the intensity changes, and the color; while the next layers capture higher-level concepts including the angle and the surface. Finally, the last layers are responsible for assigning the whole or main parts of the images to different classes.

The convolution kernels are in the form of integer matrices to preserve the affine characteristics of different transformations of the spatial perspectives including line detection, Sobel filter, gradient masks, smoothing and blurring, Laplacian filter, and arithmetic mean. The pooling layers are used to reduce the size of Convolved features while preserving the discriminations. 
The common pooling layers are the max- and the average- pooling that can be applied locally or globally \cite{zhou2016learning}. Normalization layers such as batch normalization are used in ConvNets to transform the activation of the layers to help accelerate the speed of learning and preserve convergence.

\textit{1D ConvNets} which are customized for 1 dimensional signals can offer better performance in time series, however, they still  suffered from similar issues as 2D Convents such as the need for a large set of annotated training instances \cite{kiranyaz2015real}. 
Representation learning approaches also use other elements such as restricted Boltzmann machines (RBMs) \cite{smolensky1986info}, autoencoder \cite{bourlard1988auto,hinton1994autoencoders}, sparse coding \cite{olshausen1997sparse,lee2007efficient}, and clustering methods \cite{lee2009convolutional,coates2012learning,xie2016unsupervised,9143092,ghazanfari2020deep}.

The deep-learning approaches for time series classification methods can be divided into two classes of discriminative and generative methods as mentioned in \cite{fawaz2019deep}. Generative methods use a primary pre-training with unsupervised approaches before training the classifier. The generative approaches typically are  based on the deep learning elements such as stacked denoising auto-encoders (SDAEs) \cite{bengio2013generalized}, generative CNN-based model \cite{mittelman2015time,wang2016representation}, deep belief networks (DBNs) \cite{banerjee2019deep}, and RNN auto-encoders \cite{mehdiyev2017time}. In \cite{fawaz2019deep}, the discriminative approaches are divided into two groups of feature engineering and end-to-end techniques. In feature engineering approaches, several low-level features are extracted and passed to the descriptive deep learning methods while in end-to-end approaches, even the low-level features are obtained based on feature learning. Several discriminate approaches based on feature engineering are mentioned in \cite{fawaz2019deep} based on the transformation of time series to images \cite{wang2015imaging,wang2015spatially, hatami2018classification}.

Most recent methods proposed for time series classification are mostly based on deep learning composed of ConvNets, sparse coding, RNN, and autoencoder units. Some examples include multi-channel deep convolutional neural network \cite{zheng2014time,zheng2016exploiting}, time Le-Net \cite{le2016data}, multi-scale convolutional neural network \cite{cui2016multi}, multi-layer perceptron, fully convolutional neural network, and residual network \cite{wang2017time}, encoder \cite{serra2018towards},  time convolutional neural network \cite{zhao2017convolutional}, time-warping invariant echo state network \cite{tanisaro2016time}. The authors in \cite{wang2017time} used fully convolutional neural networks (FCNs) \cite{long2015fully} using a global average pooling \cite{lin2013network} to reduce the number of parameters for time series classification. They define global average pooling in the form of a basic block of three elements: a convolutional layer of three 1D kernels, a batch normalization layer, and a ReLU activation layer. This approach cannot scale up on long-time series.

\section {Datasets Description}

ECG analysis is a vital step toward monitoring and diagnoses for cardiovascular diseases causing 30$\%$ deaths of the world \cite{lyon2018computational}. Manual annotation of ECG recordings is a tedious process and subject to human error when the human eye cannot accurately detect some complex patterns in the signal. Therefore, ECG classification is an important problem in healthcare which can take advantage of automatic signal processing and classification mechanisms.

In this paper, we evaluate the performance of the proposed layer using two publicly available ECG recording datasets including the PhysioNet Computing Cardiology Challenge in 2015 and 2017 \cite{PhysioNet15, PhysioNet17}.

\noindent \textbf{Reducing False Arrhythmia Alarms in the ICUs}:

The PhysioNet Computing in Cardiology Challenge in 2015 focused on the problem of high false alarm rates in ICUs that results in the chance of missing true life-threatening alarms by medical staff \cite{PhysioNet15,clifford2015physionet}. This issue is due to the misclassification of ECG or other collected physiological signals when the signals are highly impacted by noise or motion artifacts. An example of a normal ECG beat is shown in Figure \ref{fig:ECG_Example}. The training dataset includes 5-minutes and 5-minutes and 30 seconds recordings of two ECG leads (from lead I, II, III, aVR, aVL, aVF, or MCL) and one or more pulsatile waveforms, such as arterial blood pressure (ABP), and photoplethysmogram (PPG) signals for 750 patients right before the alarm is triggered. The dataset includes five life-threatening cardiac events of asystole (ASY), extreme bradycardia (EBR), extreme tachycardia (ET), ventricular tachycardia (VT), and ventricular flutter/fibrillation (VF) and each sample is labeled as ``True'' or ``False''. The description of this dataset is presented in Table. \ref{table:2015_summary}. The test dataset for this challenge is not available to the public. Therefore, we used the training dataset for both training and test purposes with K-fold cross-validation.

\noindent \textbf{Atrial Fibrillation (AFib) Classification from a Short Single Lead ECG Recording}: 

AFib is the most common sustained cardiac arrhythmia caused by the rapid and irregular atrial activation. This condition is the most prevalent arrhythmia leading to hospital admissions in the U.S. and it has been identified as one of the leading causes of stroke and sudden cardiac death \cite{clifford2017af,elmoaqet2017new}. Common AFib detection methods include atrial activity-based methods and ventricular response analysis-based methods \cite{clifford2017af}. Atrial activity-based methods detect AFib by analyzing the absence of P waves and the presence of fibrillatory waves. These methods can achieve good performance on clean ECG signals with a high sampling rate, but are highly vulnerable to noise contamination. Ventricular response analysis-based methods are based on the detection of RR intervals and are usually more robust against noise noting the acceptable performance of R-detection methods. However, previous studies that work on AFib classification suffer from several shortcomings as mentioned in \cite{PhysioNet17} since 1) they were only focused on the classification of normal and AFib rhythms, 2) the good detection performance was obtained on clean signals, 3) the test dataset was not separated, and 4) the number of patient instances was small. The training set provided by this challenge includes 8528 samples of short recordings for a single-lead ECG (9 seconds to 60 seconds) and the corresponding label of AFib, normal, noise, and other rhythms. Similar to the 2015 challenge, we compare our results with the best results on this challenge based on the training set since the test set is not available to the public. The description of this dataset is summarized in Table.\ref{table:2017_summary}.

\section{Related works}

In the following, we review several approaches in time series classification, false alarm reduction in ICUs, and AFib classification from a short single lead ECG recording. 
In continue, we briefly explain the related works in ECG classification as an application domain to evaluate the performance of the proposed method in analyzing ECG classification. 

Several works based on deep neural networks \cite{jun2016premature} or types of ConvNets \cite{kiranyaz2015real,acharya2017application,acharya2017automated}, or a combination of LSTM and ConvNets \cite{tan2018application}  have been proposed for ECG classification and arrhythmia detection. The proposed methods in  \cite{kiranyaz2015real,acharya2017automated,tan2018application,acharya2017application,xu2018towards,lyu2018improving} can be applied when every beat are annotated, and the type of arrhythmia for each beat should be known. In \cite{xu2018towards}, the authors used deep neural networks for both of feature learning and classification. The lack of availability of annotated long ECG recordings limits the application of these supervised feature learning methods. In \cite{ziat2017spatio}, a spatio-temporal RNN is proposed that learns the dependencies by a structured latent dynamical component and a decoder to predict the observations from the latent representations.

\subsection {\textbf{Review of Related Works for False Alarm Reduction in ICUs}:}

Here we review some of the recent works related to reducing the false alarm rate in ICUs using the 2015 PhysioNet Challenge as one of the evaluation domains considered in this paper.  There are several machine learning and rule-based methods that have been proposed in this domain \cite{PhysioNet15}, but they are still not up to expected performance as they are often impacted with a high false-positive rate when the signal is contaminated by noise and motion artifacts. More importantly, they cannot offer a reliable online detection during real-time monitoring or be general for different ECG signals. The authors in \cite{plesinger2015false} proposed a method that passes the pre-processed signals through multiple tests including the regularity test and the arrhythmia test for each type of arrhythmia. This method was ranked first place in the challenge and outperformed other methods that were based on classical machine learning or representation learning. However, it mainly relies on expert knowledge; thus,  it cannot be generalized to other arrhythmia.

In \cite{kalidas2015enhancing}, the authors presented a method that trains five SVM-based arrhythmia classifiers using different features according to the type of arrhythmia. A false-positive test is conducted after each classifier. The proposed method in \cite{couto2015suppression} also considers a level of expert knowledge in terms of the required length of the signals that need to be analyzed for each arrhythmia type. In this method, first the signal quality index (SQI) for each channel is calculated. Then, a trust assignment is applied using the SQI after comparing the QRS annotation obtained from a different channel, from which the final label is decided. 

The proposed method in \cite{fallet2015multimodal}  calculates the heart rate from the ECG, ABP, and PPG signals, along with the spectral purity index obtained from the ECG signal. The veracity of the alarm is determined based on a set of decision rules and taking into account the heart rate and the spectral purity index. This method also cannot be used for automatic detection of all arrhythmia types as it relies on the expert knowledge to determine the set of rules for each arrhythmia as well as multiple physiological signals. 

In \cite{ansari2016suppression}, the authors proposed a false alarm suppression method that uses multiple models for beat detection. The detection results are verified and summarized into fused annotation results. Finally, the outputs of the previous step are used by a rule-based decision method that determines the accuracy of the alarm based on the type of arrhythmia. This method utilizes a combination of learning and expert knowledge in different phases of the proposed approach. In \cite{lehman2018representation}, a method  to suppress the false ventricular tachycardia alarm is developed. The method performs beat detection and selects a 3-second window that contains the beat with the highest ventricular probability from the last 25 seconds of the signal. The supervised denoising auto-encoder (SDAE) then takes the FFT-transformed ECG features over the window and classify whether the alarm is true or not. It should be noted that they used the MIT-BIH dataset, where each beat is annotated to train a ventricular beat classifier.
 
The proposed method in \cite{hooman2018deep} used a deep neuroevolution method that utilized genetic algorithms and \cite{alinejad2019prediction} used neural networks for arrhythmia classification. The former method is based on using handcrafted features introduced by \cite{antink2015reducing}, which includes morphological and frequency features extracted from the ECG, ABP, and PPG signals. The latter one utilized SQI, physiological features, and features introduced in obstructive sleep apnea (OSA) detection by using hand-crafted features from the ECG, ABP, and PPG signals to train the neuroevolution method. 

The proposed approach in \cite{ghazanfari2019unsupervised} is a lightweight processing using an unsupervised representation learning to extract a few features from the clusters of the beats, but the clustering step wipes out the temporal relationships among the features.
The authors in \cite{ghazanfari2019simultaneous} proposes a representation learning approach based on Bidirectional-LSTM that operates on sequences of a set of morphological features of beats.

\subsection{\textbf{Review of Related Works for AFib Classification from a Short Single Lead ECG Recording}}

In this section, we provide a brief survey of the works related to AFib classification using short ECG recordings using the 2017 PhysioNet Computing in Cardiology Challenge dataset. 
The methods that resulted in the highest score in this challenge were mostly based on a combination of machine learning and expert knowledge. For instance, the  proposed method in \cite{hong2017encase} is based on a combination of expert features and deep neural networks. Also, the authors in \cite{ teijeiro2017arrhythmia} uses a combination of machine learning and expert knowledge.

The authors in \cite{datta2017identifying} proposed a two-level cascaded binary classifier composed of three individual classifiers using five types of features for training, which included the morphological features, AFib features, heart rate variability (HRV) features, frequency features, and statistical features. The proposed method in \cite{hong2017encase} used an ensemble of classifiers trained by three types of features extracted from five types of pre-processed ECG data (long data, short data, QRS data, center wave data, and expanded data).  The features include expert features, center wave features, and deep neural network (DNN) features, where the last one is extracted from a pre-trained deep neural network. They use deep residual convolution networks, introduced in \cite{rajpurkar2017cardiologist}, which include 34 layers by modification and adding an extra layer to learn high-level features.

In \cite{zabihi2017detection}, three pre-trained classifiers (linear and quadratic discriminant analysis, and random forest) are utilized to learn meta-level features. Then, they train another random forest classifier as the final-level classifier using the meta-level features alongside with the base-level features that were directly extracted from the ECG signal. The method proposed in \cite{teijeiro2017arrhythmia} used abductive interpretation to provide global features and per-beat features of each signal for global and sequence classifications, which use XGBoost and LSTM as the machine learning methods for the corresponding classifier. The classification results are blended using the stacking technique to get the final result. The authors further improved their method in \cite{teijeiro2018abductive}.

The proposed method in \cite{cao2019atrial} handled the variable size signals where the ECG signals were re-segmented into short samples of 9 seconds. In this method, the expert knowledge is used about which parts of the signals can better represent the behavior of the arrhythmia depending on each class. 
%$\red{which parts of the signals depend on the class can be representative about the behavior of that class.}
Then, the derived wavelet frames are applied to decompose and reconstruct the samples of different scales decomposition. Multiple fast down-sampling residual convolutional neural networks (FDResNet) consist of several convolutional and fully connected layers that are trained and coupled into a multi-scale decomposition to enhance the residual convolutional neural network (MSResNet). Each FDResNet consists of 12 layers, and the proposed MSResNet consists of three FDResNets and two additional dense layers, which make the method complex in model construction.

The authors in \cite{van2019classification} constructed a neural network with convolutional and recurrent layers that were applied on the sliding windows from the denoised ECG signal. The proposed method in \cite{zhang2019fine} constructs deep convolutional neural networks (DCNN) which take the Short-Time Fourier Transformed (STFT) of the ECG signal of different lengths as the input. An online decision fusion method will then fuse the decisions from different models into a more precise one.  In \cite{xiong2018ecg}, the authors proposed RhythmNet that includes 16 convolutional layers with 16 residual blocks, three recurrent layers, and
two fully connected layers to analyze the ECG recordings with various lengths. 

The method proposed in \cite{goodfellow2018towards} used the ECG waveforms which were normalized to the median of the R-peak amplitude as the input and constructed a deep convolutional neural network to train the model. In \cite{parvaneh2018analyzing}, the proposed method trained a dense convolutional neural network (DenseNet) with spectrogram features after filtering out noisy signals with low SQI. A post-processing step was taken if the classification result is less than a heuristic threshold for the normal rhythm and the other rhythm in probability, where 437 hand-crafted features are used in Adaboost-abstain classifier. 
The authors in \cite{warrick2018ensembling} proposed a method that uses re-segmented ECG signal as the input to ten DCNNs with a convolution layer,  a dense layer, and multiple LSTM layers. The results of each classifier are ensembled to produce the final result.

\begin{figure}[h!]
\centering
\includegraphics[width=0.95\columnwidth]{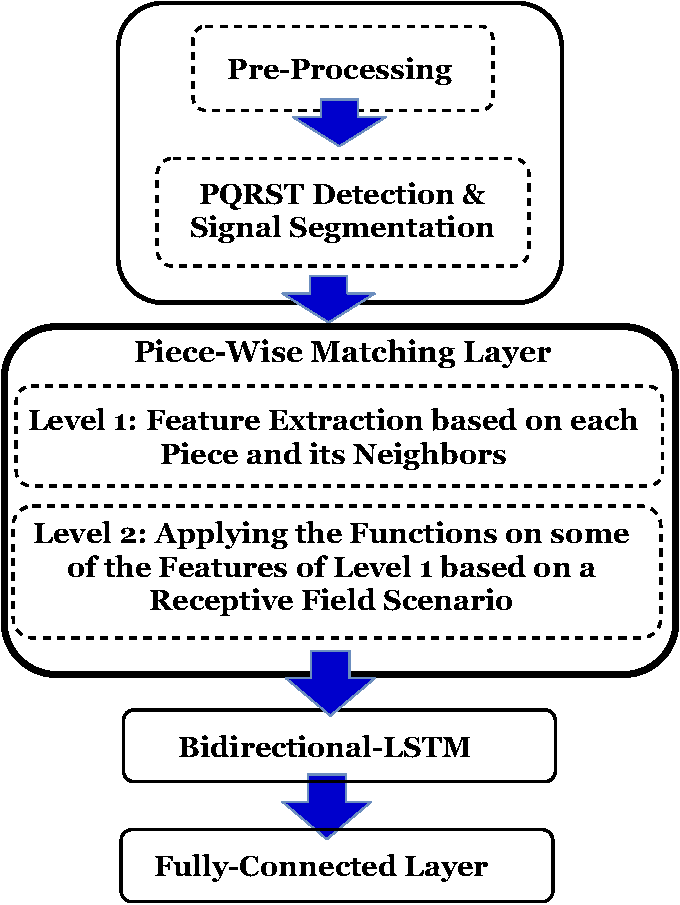}
\caption{The main parts of the proposed methods are depicted. We proposed \textit{Piece-wise matching layer} as a middle layer for the Bi-LSTM. }
\label{fig:whole_diagram_app}
\end{figure}

\section{ECG Classification based on Piece-wise Matching Layer}

The proposed approach constitutes of three phases as shown in Figure \ref{fig:whole_diagram_app}. The first phase includes three steps. We first pre-process the signal for the false alarm challenge by removing the noise and invalid parts of the signals, as described in Section \ref{subsec:processing}. Since the noise is one of the dataset's classes in the AFib challenge, we do not remove the invalid parts and noise of the signal when working with the AFib dataset. A segmentation approach based on \cite{plesinger2015false} is then applied to the pre-processed signal to extract the location of PQRST waves as described in Section \ref{subsec_segmentation}. There are two levels in the piece-wise matching layer. At the first level, a set of features are extracted from each segment as described in Tables \ref{table: morph_features}, \ref{table: stati_features}, and \ref{table:freq_features}. The features which are marked with $*$ in these tables are used in their original format without being used at the second level. At the second level, we apply some functions based on the scenario of the receptive field on some of the extracted low-level features from the signal's pieces. Both of these sets of features are fed to the bidirectional long short-term memory (Bi-LSTM). Bi-LSTM is followed by a fully connected layer to obtain the classification labels. Bi-LSTM is selected here knowing its performance in analyzing the sequence data and time series especially the long ones. The architecture of the whole structure is depicted in Figure \ref{fig:whole_diagram_app}. 

\begin{figure}[h]
\centering
\includegraphics[width=0.95\columnwidth]{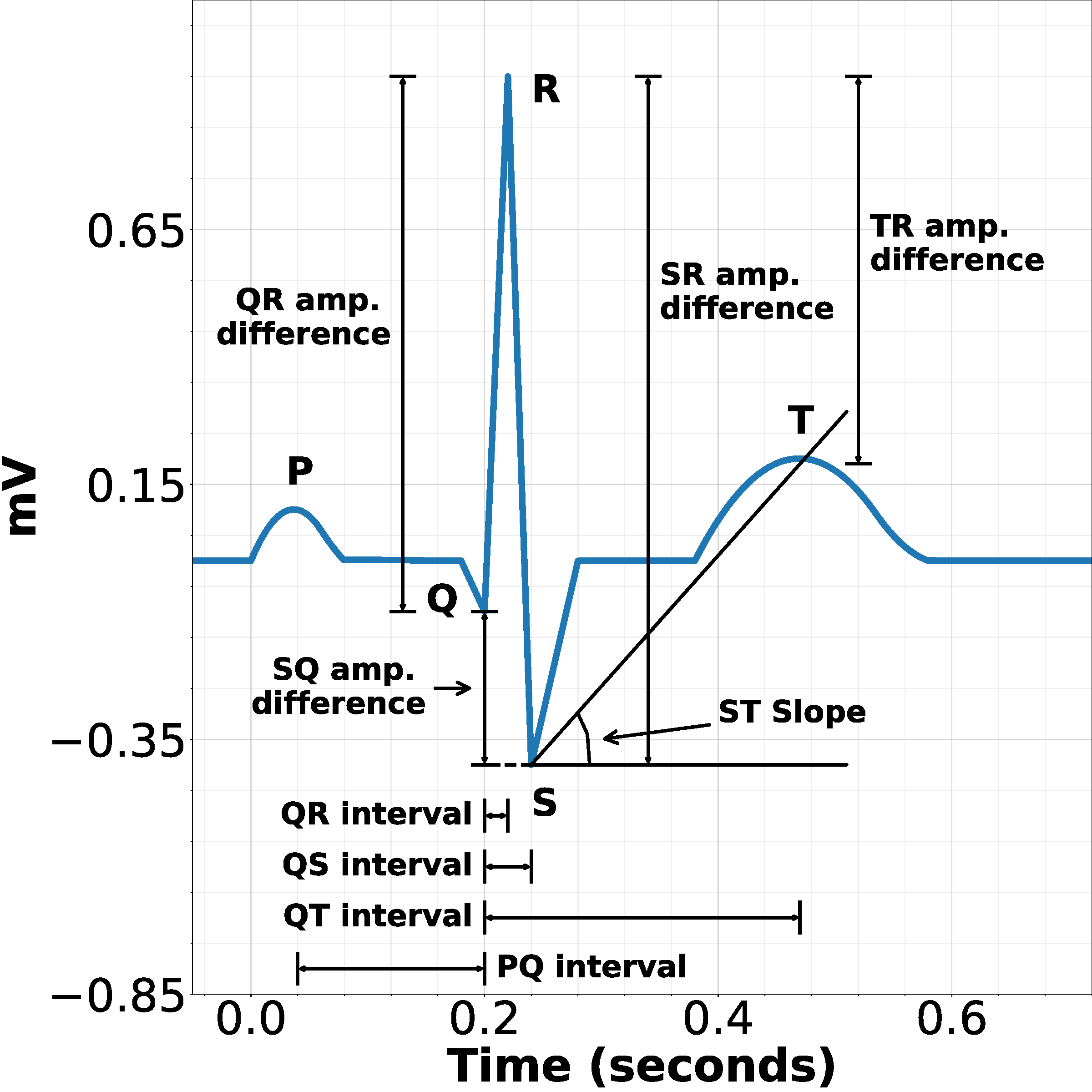}
\caption{An sample normal ECG beat. The beat as the skeleton is the quasi-periodic part through the signal that can have considerable different forms. We call that as a piece that the template or beat happens in that. Each beat has 5 elements that are shown as PQRST.}  
\label{fig:ECG_Example}
\end{figure}

\subsection{Pre-processing} \label{subsec:processing}

In the false alarm dataset, we first choose one of the two ECG signals of each record. The default choice is the ECG lead II. When the lead-II signal is not available for a patient, we select another lead in the order of I, aVF, and V. If none of these leads are available, we use the first ECG channel by default. For the training set of the 2015 PhysioNet challenge's dataset, there are 22 samples that their ECG lead II  is not available. Also, there is only one patient, ID: a675l, which does not include any of the aforementioned four leads. Hence, we use the first available ECG channel for these samples that is lead III. %Table. \ref{other_sample} in represents the list of these patients.

For the false alarm reduction challenge,  the ECG signals are first denoised by removing the sharp spikes, the flat lines, and high-frequency noises. We use the method proposed by \cite{plesinger2015false} to detect and remove invalid parts of the ECG signal. The method searches the signal inside a 2-second window and detects high-frequency noise by looking at the amplitude envelope of the signal at the frequency range of 70-90 Hz. It also detects saturated areas, which include the aforementioned the sharp spikes and the flat lines by analyzing the histogram of the 2-second window. The marked invalid parts are removed before further application. If a signal has more than 80\% region, it is marked as invalid and will be treated as noise and will be labeled as a false alarm. During the experiment, the ECG signals from 12 records are detected that have more than 80\% invalid regions and are removed from the following process. All these records turn out to be false alarm according to their true labels.

\subsection{PQRST Detection and Signal Segmentation}\label{subsec_segmentation}

We search for the QRS complex locations by using the proposed method in \cite{plesinger2015false}, which detects the R wave locations based on amplitude envelopes of the ECG signal. Figure \ref{fig:ECG_Example} shows an example of a normal ECG signal. It detects the local maxima by looking at the difference among three frequency ranges (1-8 Hz, 5-25 Hz, and 50-70 Hz) and using descriptive statistics to determine whether or not the maxima is an R wave location. Then, we refine the results and find PQST locations by searching for peaks and valleys within an interval around the R locations. If two segments are too close to each other, we compare their QRS complex amplitudes with the rest of the signal and remove the improper ones. 

\begin{table}[H]
\caption{The list of morphological features extracted per beats \cite{ghazanfari2019simultaneous}. } \label{table: morph_features}
\resizebox{1\columnwidth}{!}{
\begin{tabular}{|c|c|}
\hline
\textbf{Morphological Feature Name} & \textbf{Number of Features} \\ 
\hline\hline
PQRST amplitude &  5 \\ \cline{1-1} \cline{2-2}
PQRST time interval &  5 \\ \cline{1-1} \cline{2-2}
PQRST time interval difference &  5 \\ \cline{1-1} \cline{2-2}
RR energy & 1 \\ \cline{1-1} \cline{2-2}
Amp difference of SQ &  1 \\ \cline{1-1} \cline{2-2}
Amp ratio of SR, SR(wrt Q), TR, and QR &  4 \\ \cline{1-1} \cline{2-2}
Width difference between & \multirow{2}*{4} \\ QS, QR, QT, and PQ &   \\ \cline{1-1} \cline{2-2}
Slope between ST, QR, RS, Sx, and PQ &  5 \\ \cline{1-1} \cline{2-2}
ST neg slope, ST zero crossing point* &  2 \\ \cline{1-1} \cline{2-2}
Bazett, Fridericia, Sagie QT formula &  3 \\ \cline{1-1} \cline{2-2}
RR cluster distance* &  3 \\ \cline{1-1} \cline{2-2}

PPRR ratio & 1 \\ \cline{1-1} \cline{2-2} 
\hline
\end{tabular}}
\end{table}

\begin{table}[ht]
\caption{The list of statistical features extracted per beats. }
\label{table: stati_features}

\resizebox{1\columnwidth}{!}{
\begin{tabular}{|c|c|}
\hline
\textbf{Statistical Feature Name} & \textbf{Number of Features} \\ 
\hline\hline
PQRST wavelet entropy &  5 \\ \cline{1-1} \cline{2-2}
PQRST Hjorth parameter &  15 \\ \cline{1-1} \cline{2-2}
Shannon, Tsallis, Renyi entropy* &  15 \\ \cline{1-1} \cline{2-2}
Zero crossing ratio & 1 \\ \cline{1-1} \cline{2-2}
Linear predictive coefficients* &  11 \\ 
\cline{1-1} \cline{2-2} 
\hline
\end{tabular}
}
\end{table}
\begin{table}[H]
\centering
\caption{The list of frequency features that are extracted per beats.\label{table:freq_features}}
\resizebox{1\columnwidth}{!}{
\begin{tabular}{|c|c|}
\hline
\textbf{Frequency Feature Name} & \textbf{Number of Features} \\
\hline\hline
Spectrum centroid, roll-off, & \multirow{3}*{4}\\ kurtosis, skewness; (calculated  & \\ based on FFT )* &\\ \cline{1-1} \cline{2-2}
Short time energy, & \multirow{8}*{9} \\ frequency centroid, kurtosis, &  \\ roll-off, mode, skewness,  &\\ 80 percent energy, &\\  trimmed mean, &\\ Difference between two  & \\largest components; &  \\  (calculated based on STFT)* &\\
\cline{1-1} \cline{2-2}
Power spectrum density &  4 \\ \cline{1-1} \cline{2-2}
\hline
\end{tabular}
}
\end{table}

\section{Piece-wise Matching Layer}

In this section, we introduce the proposed piece-wise matching layer.
The main contributions of the layer are to measure the relative spatial and temporal representation of each piece and its neighbors separately as the original features in the first level. We deploy several linear and non-linear functions including mode, median, KL divergence, log energy entropy, and Shannon entropy in an inline manner for measuring the comparative features in a receptive field. Then, we use these functions to capture the similarities and differences of some of the original features over a receptive field related to each other. The mechanism of indicating the size and the placement of the receptive field provides four scenarios. The first level and the second level features are used alongside each other as a sequence data and is fed to the Bi-LSTM followed by a fully connected layer to obtain the classification label. 

\subsection{Level 1: Feature Extraction based on each Piece and its Neighbors}

%\red{ by replicating randomly of each class as the amount of  twice of instances in the largest class of arrhythmia in the false alarm dataset.} \magenta{it is not clear how you did the augmentation? did you replicate the number of instances to have an equal number of instances for each class?}\orange{For example in false alarm, I select randomly from each class as twice the number of the largest class to equal the number of two class. The instances of the smallest class are going to be several times are seen and the instances the largest class also will visited at least twice to not be forgotten. In the AFib, since there are four classes and there some classes are highly similar I tried to give the network at least some number of instances by adding the some constant numbers.} \yellow{still not clear,}

The signal is segmented by extracting PQRST waves of beats and then several types of features are extracted from the beats. The authors in \cite{datta2017identifying} extracted five types of features from the entire recording for the  AFib classification.  The features that we utilize in this work are inspired by \cite{datta2017identifying}, where the features are calculated from the basic beats of the signal while the features in \cite{datta2017identifying} are calculated based on the entire signal. Also, we just use the morphological, statistical, and frequency features of \cite{datta2017identifying} in which they are calculated per cycle. Local processing based on the beats provides a framework for a dynamic data-driven platform that is proper for real-time monitoring and decision-making. The extracted low-level features from the signals of each arrhythmia are normalized by $z$ scores for both datasets. Since the false alarm dataset is highly imbalanced, we augmented the datasets by replicating a random set of arrhythmia instances to equal the number of each class of arrhythmia. In false alarm, we select randomly from each class and replicate to reach the number of instances of each class as twice of the number of instances the largest class. There are four classes in AFib, we added instances for normal, AFib, Other Rhythm, and Noise as 3000, 4000, 5000, and 3000  respectively by replicating the instances of classes in a random manner.   

\begin{enumerate}
\item \textit{Morphological features:} Morphological features represent the behavior of an ECG signal in the time domain using the location of PQRST waves --- they are presented in Table. \ref{table: morph_features}. This set of features includes the amplitude of different waves, time intervals between consecutive PP, QQ, RR, SS, TT, differential time intervals between two consecutive intervals PP, QQ, RR, SS, TT,  the RR energy, the amplitude difference between the S and Q waves, the amplitude ratio of SR, SR (with respect to Q), TR, and QR waves, the width difference between QS, QR, QT, and PQ wave, and the slope between ST, QR, RS, Sx (slope of the period within 0.05 seconds after S wave) and PQ waves. We calculate different QT measurements using Bazett \cite{bazett1997analysis}, Fridericia \cite{moss2003introductory}, and Sagie’s QT \cite{sagie1992improved} formula as described in the appendix. We also keep track of the negative ST slope and the zero-crossing points in the ST period \cite{ghazanfari2019simultaneous}. A detailed description of these morphological features and their measurement process is presented in \cite{datta2017identifying}. %\magenta{is that what you meant?} .%\magenta{WHAT do mean by that? DO NOT design a puzzle for the reviewers!!! Either clearly describe your features or remove them}

\item \textit{Statistical features: } A set of statistical features are extracted from each segment that include wavelet entropy, Hjorth parameter, Shannon entropy, Tsallis, Renyi entropy, zero-crossing rate, and linear prediction coefficient of the segment as described in Table. \ref{table: stati_features}. The definition and measurement criteria for these features including the Hjorth parameters and entropies are presented in the appendix \ref{sec:supplementary}.

\item \textit{Frequency features: } We apply the Fast Fourier Transform (FFT)  on each segment and extract  the centroid,  roll-off, skewness, and kurtosis. Also, we use the Short-Time Fourier Transform (STFT) to calculate the short time energy, frequency centroid, roll-off, skewness, kurtosis, mode of the frequency components, and the difference between the two largest frequency components. We also extract Power spectrum densities for 0-2Hz, 2-4Hz, 4-10Hz and 10-150Hz ranges as described in Table. \ref{table:freq_features}. 
\end{enumerate}

%Also, in order to evaluate how the segment is different from or similar to other segments, we keep track of the mean, mode, and median of these feature and calculate the distance to these value for some of the morphological features and following features.

These three sets of features can describe the behavior of each beat and then we follow their behavior through the signal.

\begin{figure*}[h]
\centering

\includegraphics[width=1.95\columnwidth,height=8cm]{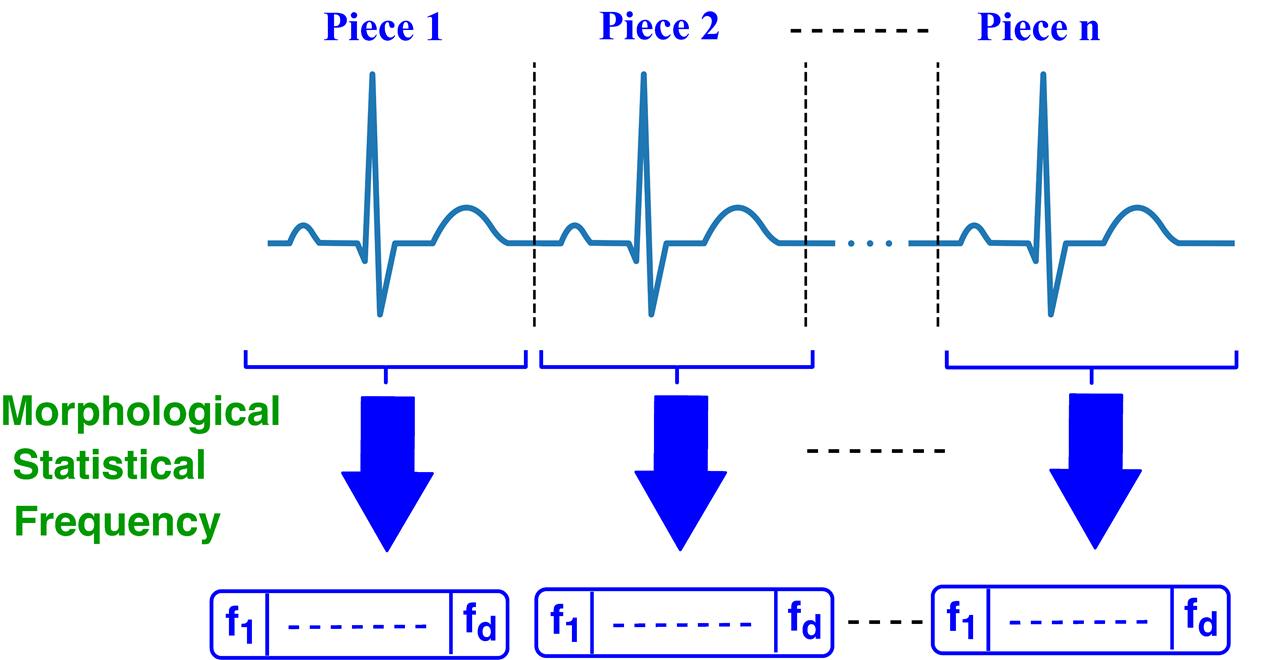}
\caption{An example of the ECG signal. The signal can be divided to several pieces (periods). Several low-level features are extracted from each piece in form of a vector. }
\label{fig:feature_extract}
%\vskip 0.25in
\end{figure*}

\begin{figure*}[h]
\centering
\includegraphics[width=1.95\columnwidth,height=12cm]{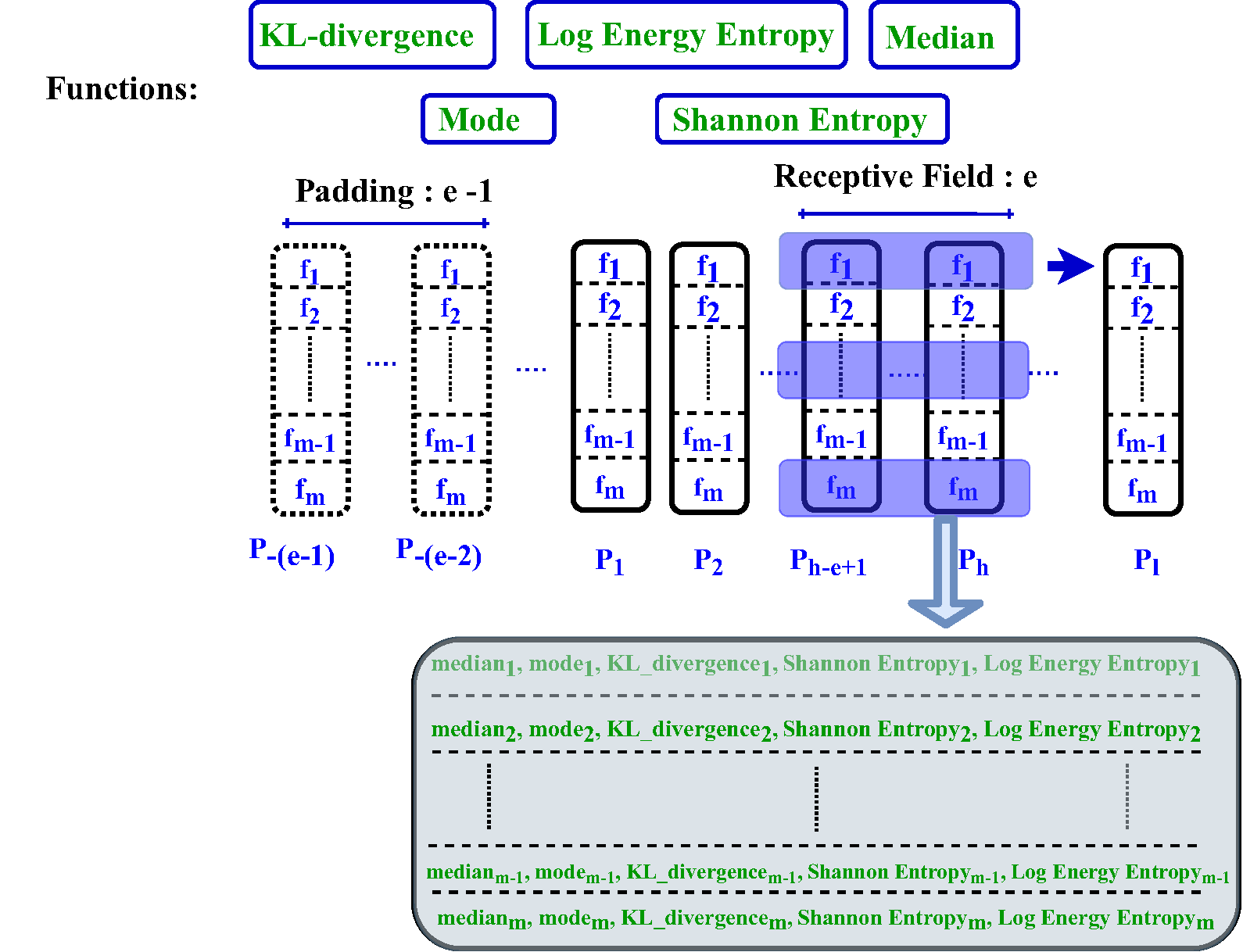} 
\caption{The $d$ receptive fields in length of $e$ that covers $d$ features from $(h-e+1)^{th}$ piece to ${h}^{th}$ piece.}
\label{fig:sliding}
\end{figure*}

\begin{figure*}[h]
\centering
\includegraphics[width=1.85\columnwidth, height=12cm]{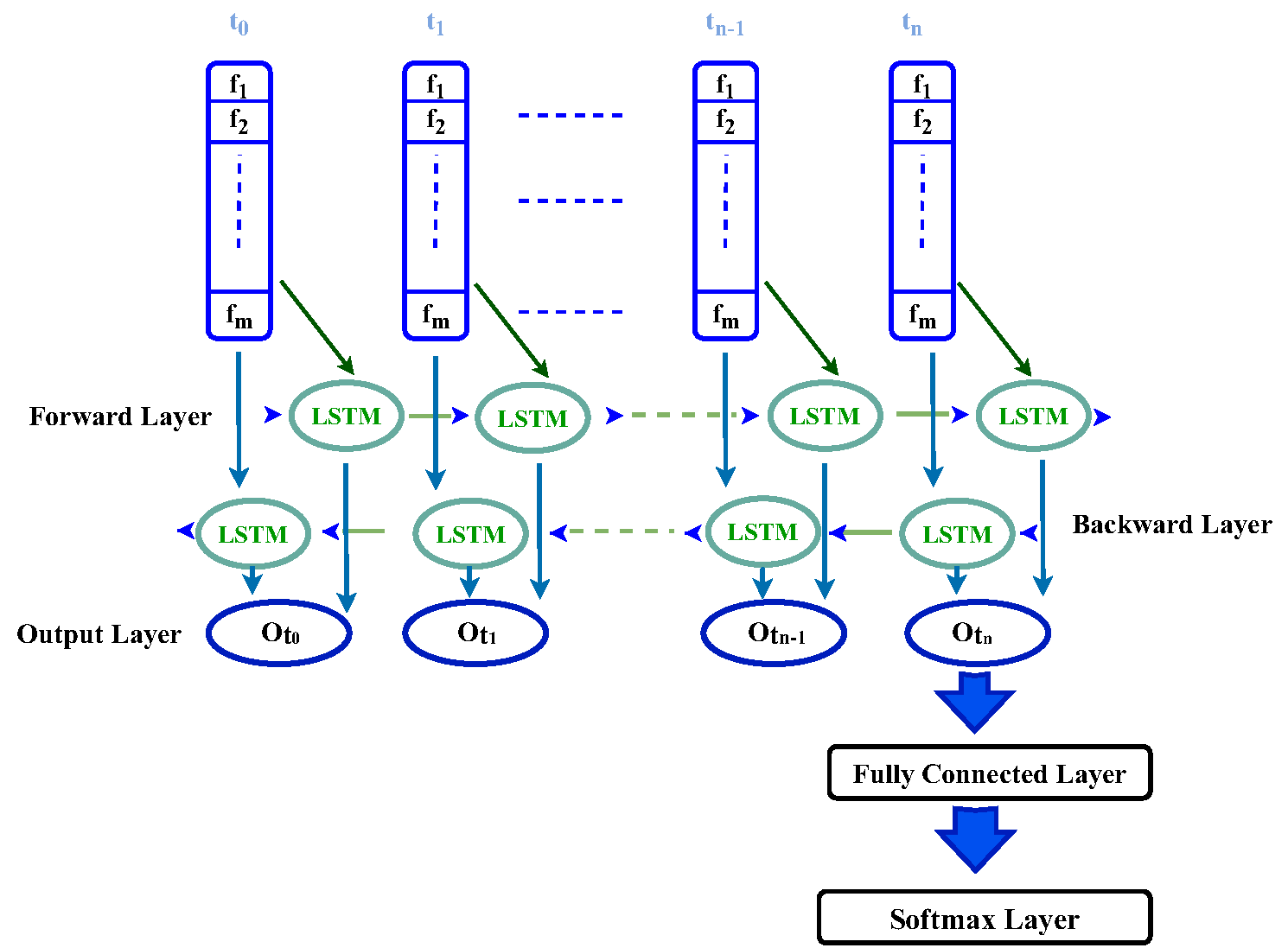} 
\caption{The features of level 1 and level 2 alongside each other per beat are fed as the features per time step to the Bi-LSTM.}
\label{fig:bilstm}
\end{figure*}

%\magenta{For your record, I edited this text before in one of your ten versions of the paper and yet again I have to revise it again!! It is a waste of my time to read the same paper over and over again and you again bring back your sloppy text. The description of the method is very confusing. It feels like you intentionally make it convoluted to make the idea look complex while it is very simple! }

%\red{, of each piece as a one dimension vector as shown in Figure.\ref{fig:feature_extract}.}
 %\red{and each feature in the vector as a dimension. In other words, a signal is transformed from one dimension, $y$ values, over $t$ to much fewer $t$ but $d$ dimensions.}
\subsection{Level 2: Applying the functions on some of the Features of Level 1 based on a Receptive Field Scenario }

\subsubsection{Pieces and features}

In the proposed piece-wise layer for the analysis of ECG signals, we consider each beat as the periodic part of the signal as a piece (e.g each heartbeat for the ECG signal --- Figure \ref{fig:ECG_Example}). The number of features extracted from each piece is denoted by $d$ including the morphological, statistical, and frequency features. Now, each signal is considered as a sequence of vectors of its pieces. In the formal description, let us denote a time series signal with $n$ time steps by $A= \{y_{1},...,y_{n}\}$. This time series signal can be divided into multiple sub-parts where each part includes a piece as a sequence of time steps, defined by $t_{c}=\{y_g,...,y_j\}$ in which $j-g+1$ is the length of that piece. A signal can be described based on its pieces as $A=\{t_{1},...,t_{l}\}$ in which $l$ is the number of pieces in the time series $A$. A schematic description of the features, pieces, and the operation is depicted in Figure \ref{fig:feature_extract}. 

%\green{As it is shown in Figure \ref{fig:sliding}, the $m$ purple windows in length of $e$ in which each one is placed on the same sequence of feature's vectors}. \magenta{the sentence is not correct}

In a brief review, the input signals are quasi-periodic; thus, the lengths of the pieces are typically variables. Therefore, we first divide the signal to its periodic pieces that each piece includes a beat (QRS) and then extract a set of low-level features from each piece as shown in Figure \ref{fig:feature_extract}. The extracted low-level features that are not marked with $*$ in Tables \ref{table: morph_features}, \ref{table: stati_features}, and \ref{table:freq_features} are fed to some functions based on a receptive field scenario.  These features, which are drawn from the original set of features of each piece, are shown by a vector $\vec{p}_{c}$ in which $\vec{p}_{c}=\{f_1,...,f_g\}$. $\vec{p}_{c}$  is the subset of all the extracted features of the piece $\{f_1,...,f_d\}$ - corresponding to the feature set of the piece $t_c$ which  are not marked with $*$. The length of  vector $\vec{p}_{c}$ is $m$, $m << d$, and is the same for all the pieces. ${p}_{c,k}$ refers to the $k^{th}$ feature, $f_k$, of $m$ features of $\vec{p}_{c}$. There are $m$ receptive fields in length of $e$ as each receptive field is placed as a one dimension window just on one type of features while its place and its mechanism are indicated depending on the scenario. For example, a receptive field ${p}_{b,k},...,{p}_{h,k}$ is placed on the $k^{th}$ feature from the $b^{th}$ piece to the $h^{th}$ piece. 
Now, we calculate the comparative temporal and spatial aspects of each feature as a unique dimension over the receptive field, as a sequence of several pieces, separately. 
The functions that are shown in Figure \ref{fig:sliding} calculate the second level features. %In other words, each feature of a receptive field corresponds to a cell.We consider each receptive field as a 1D window over $e$ consecutive features vectors.

\subsubsection{Receptive field: offline processing, incremental processing, fixed sliding receptive field, event-triggering receptive field}
%\orange{for the event triggering I think it is. Also, I have not seen in any of these works the way of processing matters while in real these things are highly important} %\red{The novel mechanisms of placing the receptive field provide other contributions and new abilities for the proposed method in the literature.} \ DTW provides the ability to select  parts of the signal dynamically as input for the representation learning and classification.
We introduce four different scenarios including offline processing, incremental processing, fixed sliding receptive field, and event-triggering receptive field based on the strategy of placing the receptive field. The novel mechanism of event-triggering receptive field for a dynamic placing of a receptive field that is based on DTW. DTW provides the ability to select  parts of the signal dynamically as input for the representation learning and classification. We describe a receptive field with two variables $RF_{start}$ and $RF_{end}$ which show the index of beginning and ending of the receptive field, respectively. 

We consider $ RF_{end} - RF_{start} > 0$ and $RF_{start}>=1$ and  $RF_{end}<=l $ in which $l$ shows the number of the pieces of the time series. $e$ refers to the length of a receptive field which is calculated as $e = RF_{end} - RF_{start}+1$. The receptive field is defined as a 1D window over the sequence of the features. Thus, we consider $d$ receptive fields, as each receptive field is applied to just one type of features through the pieces. We disregard the signals that the number of their pieces is smaller than 4. The last three scenarios of using the receptive field can be potentially used in online data processing applications such as real-time patient monitoring.  We use $RF_{(start,end),k}$ to indicate the receptive field that starts from the $start^{th}$ piece and ends in $end^{th}$ one when it is placed on $k^{th}$ feature. In continue, we describe these four scenarios.
\begin{enumerate}
    \item \textbf{Offline processing} means the entire signal is considered as the length of each receptive field --- $ RF_{start}=1, RF_{end}=l$, and $e =l$ in which $l$ is the number of pieces of the signal and the receptive field does not slide. The corresponding receptive field $\forall h \in[1,l]$ of $p_{h,k}$ is $RF_{(1,l),k}$ in offline processing.

    \item \textbf{Incremental processing} means the signal is considered as a stream and processed as received using the portion of the signal which has been received so far. Thus, we process the signal piece by piece and the length of the receptive field increases one piece at the time period.  We need some padding pieces in the beginning to process the signal to start from the first piece. Since the current time is 1 in the beginning; thus, we use 3 padding pieces and we consider $RF_{start} = -2$ and $RF_{end} = 1$. The receptive field is $\langle p_{-2,k},p_{-1,k},p_{0,k},p_{1,k} \rangle$ in the beginning. The receptive field will be $\langle p_{1,k},\dots,p_{h,k} \rangle$ after receiving $h$ heart beats, $RF_{end}=h$ and  $RF_{start}=1$, when $h > 3$. $RF_{h}$ increases as the index of the received pieces and $RF_{start}=1$ is the same; thus,  $RF_{end} = l$ in the end and $e=l$. The receptive field in online processing is $RF_{(1,h),k}$ as the corresponding receptive field for $p_{h,k}$.
   
    \item  \textbf{Fixed sliding receptive field processing} is considered as a window of several pieces to the current time, $h$. In other words, $RF_{start} = h-e+1$ to $RF_{end} = h$ in which $e$ is the length of the fixed receptive field which is a constant value. Receptive fields are shown with purple rectangles in Figure \ref{fig:sliding}. $d$ one-dimensional receptive fields with the stride of one are placed over several pieces, $e > 1$, as depicted in Figure \ref{fig:sliding}. As it is shown in this figure, there are $e-1$ padding vectors in the left of the main vectors as each receptive field is placed in the first position from $RF_{start} = -(e-2)$ that refers to $P_{-(e-2)}$ to $RF_{end} = 1$ that refers to $P_{1}$. The receptive field slides to the right until it reaches the rightest vector $P_{l}$.  For example, the number of padding vectors is $e-1$ in the first position. The stride is defined as the size of the window shifts for each time period that is equal to one. The corresponding receptive field for $p_{{h},k}$  in fixed sliding receptive field is $RF_{(h-e+1,h),k}$.  %The number of padding vectors in a receptive is denoted with $u$.or the length of the window, $e$,
    
    \item \textbf{Event-triggering receptive field} is defined based on detecting an event that is set to recognize sudden changes through time over pieces to provide a closed-loop measuring and decision making. Such a mechanism is essential for data-driven application systems. DTW is the event triggering mechanism that determines the end of the receptive field based on the similarity of the current piece to the previous one. In other words,  a receptive field is $RF_{start} = {h-e}$ and $RF_{end} = h-1$ which includes the sequence of pieces $\langle P_{h-e}, \dots, P_{h-1} \rangle$ as $ \forall \ j \in [h-e,h-1] \implies DTW (P_{j},P_{j+1}) <= event\_threshold$ and   $(P_{h-1},P_{h}) > event\_threshold$. $e$ as the length of the receptive field is a variable indicated dynamically based on the triggering event in data. The event-based triggering receptive field is a framework for dynamic data-driven usages to provide a local and dynamic perceptive processing in ECG signals. We consider event-based triggering receptive field as a tool that dynamically triggers events based on DTW for indicating the size of the receptive field on the pieces. 
\end{enumerate}

%As is shown, the receptive fields are with length $e$ and each receptive field slides on just one type of feature . \colorbox{yellow}{does not make sense at all!} \orange{why does not make sense. We track each feature separately to have a precise tracking and learning} Thus, we consider $d$ receptive fields, they are shown with purple rectangles in Figure \ref{fig:sliding}. As is shown, the receptive fields are with length $e$ and each receptive field slides on just one type of feature through the pieces with stride one.

\subsubsection{Functions, Receptive Field Features, and Bi-LSTM}

The second step is applying the proposed functions of median, mode, KL divergence, Shannon entropy, and log energy entropy based on one of the receptive field scenarios. The proposed function calculates the similarity and discrimination of temporal and spectral perspectives among the pieces of the receptive field. In the end, these comparative features are added to the original extracted features and considered as the input for the Bi-LSTM. We showed $Log Energy$ with $log En$ in the continue. We presented the formula of the functions for the fixed sliding receptive field scenario in the following:

%${median}_{p_{j,k}}$ is calculated as 
\[{median}_{p_{j,k}}= {p}_{j,k} - median({p}_{j-e+1,k},...,{p}_{j,k}).\]
%${mode}_{p_{j,k}}$ is calculated as 
\[{mode}_{p_{j,k}}= {p}_{j,k} - mode({p}_{j-e+1,k},...,{p}_{j,k}).\]
%${Shannon\ entropy}_{p_{j,k}}$ is calculated as 
\[{Shannon\ Entropy}_{p_{j,k}}=Shannon\  Entropy({p}_{j-e+1,k},...,{p}_{j,k}).\]
%${log\ on\ entropy}_{p_{j,k}}$ is calculated as
\[{Log\ En\ Entropy}_{p_{j,k}}=\\
Log\ En\ Entropy ({p}_{j-e+1,k},...,{p}_{j,k}).\]
%${KL}_{p_{j,k}}$ is calculated as
\[{KL}_{p_{j,k}}= KL\_function({p}_{j,k} , mean({p}_{j-e+1,k},...,{p}_{j,k})).\]

Finally, the new set of features per piece will be as follow:
\begin{equation} \label{eq1}
\begin{split}
\langle {p_{j,k}}, {median}_{p_{j,k}},{mode}_{p_{j,k}},{Shannon\ entropy}_{p_{j,k}}\\, {log\ energy\ entropy}_{p_{j,k}}, {KL}_{p_{j,k}} \rangle
\end{split}
\end{equation}

%\red{alongside of the original ones form the new set of the features per time}.

In order to learn the relationships among these two-level features with each other and through the time with the labels, a machine learning approach is required that captures the time dependencies among the sequence of features. Bidirectional recurrent neural networks (BRNN)  \cite{schuster1997bidirectional} is the extension of recurrent networks that take advantage of the past to future and future to the past information. Bi-LSTM is built based on the idea of forward and backward layers on the LSTM architecture. Bi-LSTM is a known machine learning method for sequence data and time series especially the long ones. Thereby, the features are fed to Bi-LSTM as a version of recurrent neural networks that have memories to capture the time dependencies as shown in Figure \ref{fig:bilstm}. Bi-LSTM is the extension of LSTM that takes advantage of another layer called  backward layer that captures information from the future to that past. Thus, Bi-LSTM utilizes two LSTMs that use both of the past and future information in the receptive field \cite{schuster1997bidirectional}. The forward layer handles the positive time direction and the backward ones are responsible for the negative direction from the future to the past. These two layers are not connected to each other.

Finally, the Bi-LSTM block is followed by a fully connected layer to obtain the classification label. We follow \cite{ghazanfari2019simultaneous} by considering the input layer as the number of features per each piece for Bi-LSTM as is shown in Figure \ref{fig:bilstm}. The learning of the network is based on adam \cite{kingma2014adam}. 
 
\begin{table*}[t]
\caption{Comparison of TPR, TNR, and the challenge score of recent false alarm reduction methods using PhysioNet 2015 Challenge dataset. The table is partially presented in \cite{ghazanfari2019simultaneous}.}
\begin{center}
\resizebox{2\columnwidth}{!}{
\begin{tabular}{|c|c|c|c|c|c|c|}
\hline
 \textbf{Method} & \textbf{Features} & \textbf{Input Signal} & \textbf{TPR}  & \textbf{TNR}  & \textbf{Score} \\ \hline
Rule-based Arrhythmia Test \cite{plesinger2015false} & Hand crafted features & ECG, ABP, PPG & 93.5 & 86.0 & 80.8\\\hline
SVM-based Classifier \cite{kalidas2015enhancing} & Time and frequency features & ECG, ABP, PPG & 85 & 93.2 & 72.9\\\hline
Trust Assignment and Thresholding \cite{couto2015suppression}$^1$ & SQI and SPI & ECG, ABP, PPG & 89.0 & 91.0 & 79.0 \\\hline
Feature-based Decision Making \cite{fallet2015multimodal} & Heart rate and SPI & ECG, ABP, PPG & 93.5 & 77.9 & 76.3\\\hline
Decision Tree and Rule-based \cite{ansari2016suppression}& Detected beats & ECG, ABP, PPG & 97.0 & 92.0 & 89.1 \\ \hline
%Supervised denoising autoencoder (SDAE) \cite{lehman2018representation}$^3$  & FFT of beats & ECG & 89 & 86 & 77.6 \\
\cline{4-5}
Deep Neuro-evolution \cite{hooman2018deep}$^2$  & Time and frequency features & ECG lead II, V, PPG &
\multicolumn{2}{c|}{91.9} & 86.8\\\hline
\cline{4-5}
Neural Network \cite{alinejad2019prediction}  & SQI, physiological, and OSA features & ECG lead II & 81.6 & 85.2 & 80.6 \\\hline
Unsupervised Feature Learning \cite{ghazanfari2019unsupervised}  & Morphological features & ECG lead II & 81 & 83 & -- \\\hline
Simultaneous multiple feature tracking \cite{ghazanfari2019simultaneous} & Morphological features & ECG lead II & 97.3 & 95.5 & 90.8\\
\hline\hline
Tree$^3$ & Wavelet features & ECG lead II& 65.3 & 80.2 & 52.8 \\\hline
%                           & Coarse tree & Wavelet features & ECG & 36.7 & 93.6 & 40.9 \\
Linear Discriminant & Wavelet features  & ECG lead II & 65.0 & 76.1 & 50.8 \\\hline
Logistic Regression &  Wavelet features & ECG lead II & 55.4 & 67.8 & 41.3\\\hline
Naive Bayes$^4$ & Wavelet features  & ECG lead II & 52.7 & 76.1 & 43.0 \\\hline
SVM$^5$ &   Wavelet features & ECG lead II & 55.1 & 89.0 & 49.6 \\\hline
%& Coarse KNN & Wavelet features & ECG & 60.2 & 77.2 & 48.1 \\
KNN$^6$ & Wavelet features  & ECG lead II & 68.0 & 71.1 & 50.8 \\\hline
Boosted Tree & Wavelet features& ECG lead II& 69.4 & 88.4 & 59.5 \\\hline
Ensemble$^{7}$ & Wavelet features  & ECG lead II & 76.2 & 82.9 & 62.7 \\\hline
PCA (Quadratic Discriminant)$^{8}$ &Wavelet features & ECG lead II & 98.6 & 9.0 & 43.2 \\
\hline
\hline
\textbf{Piece-wise Matching Layer - Scenario Offline} & Local features per beats & ECG lead II &  97.52 & 98.69 & 92.57 \\\hline
\textbf{Piece-wise Matching Layer - Scenario Online} & Local features per beats & ECG lead II &  98.2 & 97.71 & 93.99 \\\hline
\textbf{Piece-wise Matching Layer - Scenario Fixed} & Local features per beats & ECG lead II &  97.76 & 98.37 & 93.06 \\\hline
\textbf{Piece-wise Matching Layer - Scenario Event} & Local features per beats & ECG lead II &  \textbf{98.42} & \textbf{98.37} & \textbf{94.85} \\\hline
\end{tabular}}
\end{center}

$^1$ The reported results in \cite{couto2015suppression} are based on training on the public training dataset and testing on the private test  dataset from PhysioNet 2015 Challenge.

$^2$ The reported results in \cite{hooman2018deep} are not provided TPR and TNR, so the accuracy is reported instead.  The result is based on a subset (572 records) of PhysioNet 2015 Challenge public dataset.

We experimented various methods for each baseline approach in MATLAB Classification Learner and reported the best results for that version. The specific name of the  method we used are listed as follow: $^3$Medium Tree.  $^4$Kernel Naive Bayes. $^5$Fine Gaussian SVM. $^6$Cubic KNN. $^{7}$RUSBoosted Tree. 

$^{8}$ The result scores of different classifiers utilizing principle component analysis (with 95, 98, and 99 percents explained variance) are far lower than those without it. Thus, we just report the best result of the baseline using PCA with 99 percents explained variance.

\label{2015_performance}
\end{table*}

\begin{table*}[t]
\caption{ Comparison of TPR, TNR result per arrhythmia type using PhysioNet 2015 Challenge public dataset. The table is partially presented in \cite{ghazanfari2019simultaneous}.}
\begin{center}
\resizebox{2\columnwidth}{!}{
\begin{tabular}{|c|c|c|c|c|c|c|c|c|c|c|}
\hline
\textbf{Method} & \multicolumn{2}{c|}{\textbf{ASY1}} & \multicolumn{2}{c|}{\textbf{EBR2}} & \multicolumn{2}{c|}{\textbf{ETC3}} & \multicolumn{2}{c|}{\textbf{VFB4}} & \multicolumn{2}{c|}{\textbf{VTA5}}\\
\cline{2-11}
& \textbf{TPR} & \textbf{TNR} & \textbf{TPR} & \textbf{TNR}  & \textbf{TPR}  & \textbf{TNR}  & \textbf{TPR}  & \textbf{TNR}  & \textbf{TPR}  & \textbf{TNR} \\
\hline\hline

Rule-based Arrhythmia Test\cite{plesinger2015false} & 
95.5 & 92.0 & 97.8 & 74.4 &  99.2 & 88.9 & 83.3 & 100.0 & 83.1 & 82.5\\\hline
SVM-based Classifier \cite{kalidas2015enhancing} & 
77.3 & 93.0 & 100.0 & 93.0 & 100.0 & 66.7 & 16.7 & 96.2 & 61.8 & 93.7\\\hline
Trust Assignment and Thresholding \cite{couto2015suppression}$^1$ & 
78 & 94 & 95 & 66 & 100 & 80 & 89 & 96 & 69 & 95 \\\hline
Feature-based Decision Making\cite{fallet2015multimodal} & 
100.0 & 88.0 & 97.8 & 62.8 & 96.9 & 33.3 & 83.3 & 84.6 & 85.4 & 76.6\\\hline
Decision Tree and Rule-based \cite{ansari2016suppression}& 95 & 86 & 98 & 88 & 98 & 67 & 50 & 100 & 97 & 94\\\hline

% Hooman et al. \cite{hooman2018deep}$^3$ & Neural network \& genetic algorithm & & & & & & & & & &\\
Neural Network \cite{alinejad2019prediction} & 83 & 93 & 73 & 50 & 100 & 100 & 100 & 100 & 52 & 83 \\\hline
Supervised denoising autoencoder (SDAE)\cite{lehman2018representation}$^2$ & -& -& & -& -&-& -& -training & 89.0 & 86.0  \\\hline
Simultaneous multiple feature tracking \cite{ghazanfari2019simultaneous} & 98.97 &87.5 & 100& 95.62 &100 &99.24 & 100& 100& 95.7& 91.76\\ \hline
\hline
Tree$^3$ &	45.5	& 85.0	& 54.3	& 69.8	& 91.6	& 0		& 16.7	& 80.8	& 51.7	& 81.0\\\hline
%        & Coarse tree &  		36.4	& 90.0	& 52.2	& 65.1	& 92.4	& 0		& 16.7	& 80.8	& 55.1	& 84.5\\
Linear Discriminant & 	72.7	& 63.0	& 69.6	& 58.1	& 77.9	& 22.2	& 83.3	& 80.8	& 48.3	& 73.4\\\hline
Logistic Regression &  68.2	& 58.0	& 58.7	& 48.8	& 79.4	& 22.2	& 33.3	& 69.2	& 50.6	& 68.7\\\hline
Naive Bayes$^4$         & 	9.1		& 93.0	& 91.3	& 30.2	& 96.9	& 0		& 0		& 94.2	& 41.6	& 85.7\\\hline
SVM$^5$                 & 	13.6	& 99.0	& 87.0	& 34.9	& 100.0	& 0		& 0		& 100.0	& 23.6	& 96.0\\\hline
%        & Coarse KNN &  			0		& 100.0	& 100.0	& 0		& 100.0 & 0		& 0 	& 100.0	& 1.1	& 99.6\\
KNN$^6$                 &			9.1		& 96.0	& 76.1	& 55.8	& 100.0 & 0		& 0		& 100.0	& 47.2	& 90.1\\\hline
%        & Boosted Tree & 		22.7	& 89.0	& 100.0	& 0		& 100.0 & 0		& 0		& 100.0	& 50.6	& 88.1\\
Ensemble$^{7}$           &		72.7	& 73.0	& 67.4	& 51.2	& 70.2	& 33.3	& 50.0	& 73.1	& 69.7	& 77.8\\\hline
\hline
\textbf{Piece-wise Matching Layer - Scenario Offline}  &		100.0	& 96.15	& 97.67	& 97.83	& 88.89	& 99.24	& 100.0	& 100.0	& 96.33	& 98.96\\\hline

\textbf{Piece-wise Matching Layer - Scenario Online}  &		98.96	& 92.31	& 97.67	& 97.83	& 88.89	& 99.24	& 100.0	& 100.0	& 97.96	& 96.88\\\hline

\textbf{Piece-wise Matching Layer - Scenario Fixed}  &		98.96	& 92.31	& 97.67	& 100.0	& 88.89	& 99.24	& 100.0	& 100.0	& 97.14	& 97.92\\\hline

\textbf{Piece-wise Matching Layer - Scenario Event}  &		100.0	& 96.33	& 98.86	& 97.83	& 100.0	& 99.24	& 100.0	& 100.0	& 97.55	& 97.92\\\hline

\end{tabular}
}
\end{center}
%Table \ref{2015_type}.

$^1$ This result comes from \cite{couto2015suppression} and is based on the hidden dataset from PhysioNet 2015 Challenge.

$^2$ This result comes from \cite{lehman2018representation} with the task of reducing false VTA alarm. It is based on two ECG leads of 562 VTA records from PhysioNet 2015 Challenge public and hidden datasets.

We experimented various methods for each baseline category in MATLAB Classification Learner and reported the best results for that category. The specific name of the  method we used are listed as follow: $^3$Medium Tree, $^4$Kernel Naive Bayes, $^5$Fine Gaussian SVM, $^6$Cubic KNN, $^{7}$RUSBoosted Tree.

\label{2015_type}
\end{table*}

\begin{table*}
\caption{The approaches that are based on deep learning for the datasets \cite{clifford2017af,elmoaqet2017new}.}
\resizebox{2\columnwidth}{!}{
\begin{tabular}{|c|c|c|c|c|c|c|}
\hline
\textbf{The Deep Learning Approaches} & \textbf{Layer Types} & \textbf{Number of Layers} \\
\hline\hline
%Didn't use NN\cite{datta2017identifying} & & \\
Deep feature extractor \cite{hong2017encase} & Convolution, LSTM, and dense & % 13+1
14 \\\hline
%Didn't use NN\cite{zabihi2017detection} & & \\
Sequence classification \cite{teijeiro2017arrhythmia, teijeiro2018abductive} & LSTM and pooling & 8 \\\hline
FDResNet and MSResNet\cite{cao2019atrial} & Convolution, max pooling, flatten, and dense &% 10*3 + 2
32\\\hline
\cite{van2019classification} & Convolution, pooling, and LSTM & 8\\\hline
DCNNs and decision fusion \cite{zhang2019fine} & Convolution and max pooling & 4\\\hline
\cite{goodfellow2018towards} & Convolution and max pooling & 13 \\\hline
\cite{parvaneh2018analyzing} & Convolution and dense & 3\\\hline
\cite{xiong2018ecg} & Convolution, max pooling, recurrent, and dense & % 16 + 5
21\\\hline
\cite{warrick2018ensembling} & Convolution, LSTM, and max pooling & % 6*10 + ?
60 \\
\hline\hline
\textbf{Piece-wise matching Layer}  & Bi-LSTM, and dense & % 6*10 + ?
2 \\
\hline
\end{tabular}
}
\label{deep_learning_afib}
\end{table*}

\begin{table*}[h]
\caption{ Comparison of F1 score result per class and the challenge Score of recent AFib classification methods using PhysioNet 2017 Challenge public dataset \cite{clifford2017af}.}
\resizebox{2\columnwidth}{!}{
\begin{tabular}{|c|c|c|c|c|c|c|}\hline
\textbf{Method} & \textbf{Features} & \multicolumn{4}{c|}{\textbf{F1 Score}} &{|}\\
\cline{3-7}
& & \textbf{Normal} & \textbf{AF} & \textbf{Other} & \textbf{Noise} & \textbf{Score} \\
\hline\hline
Cascaded classifier(Adaboost)\cite{datta2017identifying}  & Morphological, Statistical, Frequency, HRV, and AFib & 92.0 & 82.0 & 75.0 &--& 83.0 \\\hline
Ensemble classifier(xgboost)\cite{hong2017encase} & Expert, pre-trained DNN, and centerwave features & 92.0 & 84.0 & 74.0 &--& 83.0 \\\hline
Random forest \cite{zabihi2017detection} & Multi-level features from pre-trained classifiers & 90.87 & 83.51 & 73.41 & 50.42 & 83.0 \\\hline
XGBoost \& LSTM \cite{teijeiro2017arrhythmia} & Global and per-beat features from abductive interpretation & -- & -- & -- & -- & 83.0 \\\hline
XGBoost \& LSTM \cite{teijeiro2018abductive} & Global and per-beat features from abductive interpretation & 95.3 & 83.8 & 85.0 & 71.1 & 88.0 \\\hline
MSResNet \cite{cao2019atrial} & Short segment and wavelet decomposition & 88.1 & 96.6 & 85.1 &--& 89.9 \\\hline
Convolution and recurrent neural network \cite{van2019classification} & Sliding windows from ECG & 95.09 & 92.32 & 87.28 & 92.64 & 91.56 \\\hline
% DCNNs and decision fusion \cite{zhang2019fine} & STFT ECG & & & & & 	0.996 \textcolor{blue}{?}\\
DCNN \cite{goodfellow2018towards} & Normalized ECG & 92.0 & 81.0 & 80.0 &--& 84.0 \\\hline
DenseNet \& AdaBoost-abstain \cite{parvaneh2018analyzing} & SQI, spectrogram \& hand-craft feature & 91.0 & 89.0 & 78.0 &--& 86.0 \\\hline
Convolutional recurrent neural network\cite{xiong2018ecg} & Raw ECG of 5 seconds segment & 91.9 & 85.8 & 81.6 &--& 86.4 \\\hline
Esembled classifier (ConvNets and LSTMs)\cite{warrick2018ensembling} & Re-segmented ECG & 91.0 & 81.0 & 78.0 & 48.0 & 85.3 \\\hline
\hline
Tree$^1$ & Wavelet features & 77.4 & 14.8 & 24.2 & 36.4 & 38.8 \\\hline
Linear Discriminant & Wavelet features & 76.4 & 22.8 & 28.9 &45.0 & 42.7 \\\hline
Naive Bayes$^2$ & Wavelet features & 66.8 & 26.5 & 31.5 & 21.5 & 41.6 \\\hline
SVM$^3$ &  Wavelet features & 76.8 & 19.3 & 37.2 & 48.2 & 44.4\\\hline
KNN$^4$ & Wavelet features & 75.3 & 16.8 & 30.6 & 27.4 & 40.9 \\\hline
Ensemble$^5$ & Wavelet features & 77.2 & 19.3 & 36.8 & 42.1 & 44.4 \\\hline
PCA (RUSBoosted Trees)$^6$ & Wavelet features & 50.8 & 17.7 & 34.6 & 26.5 & 34.3 \\
\hline\hline
\textbf{Piece-wise Matching Layer - Scenario Offline} & Local features per beats & 99.29 & 98.31 & 98.35 & 96.13 & 98.65 \\\hline
\textbf{Piece-wise Matching Layer - Scenario Online} & Local features per beats & 99.12 & 97.97 &  97.89 & 95.44 & 98.33 \\
\hline
\textbf{Piece-wise Matching Layer - Scenario Fixed} & Local features per beats & \textbf{99.42} & \textbf{98.38} & \textbf{98.53} & 96.99 & \textbf{98.78} \\
\hline
\textbf{Piece-wise Matching Layer - Scenario Event} & Local features per beats & 99.41 & 98.24 &  98.37 & \textbf{97.19} & 98.67 \\
\hline
\end{tabular}
}\\
We experimented various methods for each baseline category in MATLAB Classification Learner and reported the best results for that category. The specific name of the  method we used are listed below:\\
$^1$Medium Tree, $^2$Kernel Naive Bayes, $^3$Quadratic SVM, $^4$Cubic KNN, $^5$Bagged Tree.

$^6$ The result scores of baselines utilizing principle component analysis (with 95, 98, and 99 percents explained variance) are far lower than those without it. Thus, we just report the best result of the baseline using PCA with 99 percents explained variance.
\label{2017_performance}
\end{table*}

\section{Experimental Results} 

In this section, we evaluate the performance of the proposed method using two  publicly available ECG recording datasets that include a relatively small number of instances with variable length and are highly imbalanced. The results of our method are compared against the top-ranking entries from PhysioNet 2015 and 2017 Challenges \cite{clifford2015physionet,clifford2017af} as well as several recently reported approaches to compare our results  with various techniques from expert-based rules \cite{couto2015suppression,plesinger2015false}, machine learning \cite{fallet2015multimodal}, representation learning \cite{lehman2018representation,hooman2018deep,alinejad2019prediction}, and methods based on combing the rule- and machine learning-based techniques \cite{ansari2016suppression}. Various baseline classifiers are also included in the comparison.  We like to point out that the majority of commonly-used datasets in the literature of classification of time series are not long enough (i.e., less than 2000 time steps \cite{wang2017time}) and generally not periodic. However, the 2015 PhyioNet challenge dataset \cite{PhysioNet15} includes around 70 to 80 thousand time steps with a small number of instances  and the 2017 PhyioNet challenge dataset includes between 2 thousand to 15 thousand time steps \cite{PhysioNet17}. In \cite{PhysioNet17}, there are four classes including normal, AFib, Noise, and other Rhythm. Our proposed method is a general solution that is designed to handle ECG signals ranging from short to long ones as it learns the nuance patterns among beats, the periodic parts, of the signal associated with different classes to accurately classify them.% \red{while the highly noise }% \red{ and motion artifact contaminated signals.}   \red{There are the instances from different classes in \cite{PhysioNet17} like other normal, other rhythms, and AFib that there are some instances belong to them that are very similar to each other. } \textcolor{red}{The challenge is difficult in real periodic long time series since an approach should learn which nuance among the periodic parts and in where lead to each class while there are distortion and noises. } \magenta{remove all these sentences}

%\textcolor{magenta}{move to introduction} \textcolor{green}{The deep learning methods in time series classification mostly use annotated time series or the short ones, need considerable training instances, or some in advanced knowledge in form of which parts of the signal are representative of each label.} %It should be noted, there are datasets in MIT BIH \cite{} that beats are independently labelled by several cardiologists or in datasets that the length of signals are short that have not used in this paper. We want to show how the proposes layer works in confront of especially long ones and when the number of training patterns are few that are the issues of the current machine learning and deep learning ones. The challenge is difficult in real periodic long time series since an approach should learn which nuance among the periodic parts and in where lead to each class while there are distortion and noises. 

%\yellow{you need to metion that you have augmented the dataset} \yellow{therefore, it is not fair to comapre the results with } \yellow{these papers unless they have used the SAME augmentation method} 
We like to note that the test datasets for 2015 and 2017 PhysioNet Challenges are not publicly available. Therefore,  we re-implemented most aforementioned approaches on the public training datasets to report fair results. For those methods that we were not able to implement, we listed their reported results. The public training sets of these challenges are partitioned to the training and test sets based on K-fold cross validation. We used Bi-LSTM of Deep Learning toolbox of MATLAB. The Bi-LSTM is one layer that the number of its hidden nodes are equal with the number of inputs. In the false alarm challenge, the batch size is considered 32, the learning rate is $2.5*10^{-4}$, and L2 Regularization is $1.0*10^{-6}$. In the AFib challenge, the batch size is considered 16, the learning rate is $1*10^{-3}$, and L2 Regularization is $1.0*10^{-7}$.

We also report the performance of several baseline classifiers applied on wavelet features. We use discrete wavelet transform (DWT) on the entire ECG recording in which a 6-level Daubechies 8 (db8) wavelet is used to calculate the wavelet features. We used baseline category classifiers in MATLAB Classification Learner including the decision tree, linear discriminant, logistic regression, Naive Bayes, SVM, KNN, and Ensemble learning and reported results of their best variant  for each category.

%\begin{table}
%\begin{center}
%\caption{Details of records that because of many invalid parts are considered as False Alarm \colorbox{yellow}{Remove this table, only mention these samples in the evaluation section}. \label{removed_records}}
%\resizebox{1\columnwidth}{!}{
%\begin{tabular}{|c|c|c|}
%\hline
%Alarm Type & Sample ID & Total \\
%\hline
%ASY & a382s, a391l, a608s, a668s & 4 \\
%VFB & f530s & 1 \\
%VTA & v244s, v400s, v405l, v433l, v491l, v623l, v774s & 7 \\
%\hline
%\end{tabular}
%}
%\end{center}
%\end{table}

\noindent \textbf{Results on 2015 PhysioNet computing in Cardiology Dataset}: Current works in false alarm reduction can be divided to rule-based methods based on human knowledge \cite{couto2015suppression,plesinger2015false}, classical machine learning (ML) methods \cite{fallet2015multimodal}, and representation learning ones \cite{lehman2018representation,hooman2018deep,alinejad2019prediction}. The methods based on human knowledge or combinations of ML and expert knowledge methods lead to considerably better results compared to the ones only using ML  \cite{clifford2016false}, since the ML-based approaches cannot show good performance when dealing with imbalanced datasets where there are only a few instances available for some arrhythmia. Another fact we like to point out is that several examples of the best-reported results only operate on the last portion of the ECG signal based on prior knowledge that the alarm has happened in the last 10-second.  However, such knowledge is  obviously  not available in real-time patient monitoring, therefore it is very likely that the performance of such methods will be considerably degraded when dealing with long signals.  One of the key advantages of the proposed method is handling long ECG signals without any prior knowledge on the possible time of the placement of arrhythmia while the some of them  require this information.

The instances that have many invalid parts that are considered as False Alarm per alarm type are a382s, a391l, a608s, a668s in ASY, f530s in VFB, and v244s, v400s, v405l, v433l, v491l, v623l, v774s in VTA.  Table.  \ref{2015_performance} compares the performance of our method with the state-of-the-art reported techniques for all the alarm types in terms of three measures of (i) true positive rate (TPR), also known as sensitivity, (ii) true negative rate (TNR),  also called as specificity, and (iii) the challenge score calculated as 
\begin{equation}
Score =\frac{100 \cdot ( TP + TN )}{( TP + TN + FP + 5\cdot FN )}.
\label{2015_challenge_score}
\end{equation}

Table. \ref{2015_type} reports the performance for each specific arrhythmia type. It can be seen the proposed approach leads to a state of the art with a considerable margin. These results are obtained while we just used one lead of information but the most approaches are based on three leads \cite{hooman2018deep,ansari2016suppression,fallet2015multimodal,couto2015suppression, kalidas2015enhancing, plesinger2015false}, we do not use expert knowledge about the rules while some approaches are based on that \cite{couto2015suppression,plesinger2015false,ansari2016suppression}, and we do not use the expert knowledge about the possible place of arrhythmia while some of the methods rely on this information \cite{hooman2018deep,ansari2016suppression,fallet2015multimodal,couto2015suppression, kalidas2015enhancing, plesinger2015false}.

%\red{The approaches cover a variety of the known techniques while they are customized for the AFib dataset --- they are described in related works. } \orange{do not you think we shold emphasis on that these appraoches are customized considerably for that competition} \yellow{should be removed}

\noindent \textbf{Results on 2017 PhysioNet computing in Cardiology Dataset}:
 We show in Table. \ref{deep_learning_afib}, the architecture of known approaches that are based on deep learning for the AFib dataset in comparison with  the proposed approach.  Table. \ref{2017_performance} reports the comparison results based on the competition score that is the average of F1 score of three classes, normal, AF rhythm, and other rhythm, of four classes \cite{PhysioNet17}. 

\begin{equation}
Score =\frac{(F1_{Normal}+F1_{AF}+F1_{Other})}{3}
\label{2017_challenge_score}
\end{equation}

%\yellow{I still thing the red ones need to be removed} \red{and is the same in both of those challenges}  \red{ in while leads to the state of the art in both of these challenges}It can be seen in both of the competitions, all of the four scenarios lead to considerably better results than other approaches. Also, event triggering receptive field works in False Alarm and the noise class of AFib that shows its dynamic position is a strong tactic in dealing with noisy data that leads to the best reported among the proposed scenarios. The fixed receptive field provides powerful aspects in both competitions mostly in which the signals have not a considerable noise like the three classes of AFib challenge.

In this paper, the proposed layer boosts a representation learning method as a generic framework that stands on learning from scratch. The current deep learning methods require considerable training several layer networks, and their networks for long time series require tuning in the architecture to keep the efficiency for different patterns in ECG signals. The proposed approach mitigates the mentioned challenges. In fact, it shows the larger view localization based on DTW to obtain the higher level features as a powerful tool for representation learning approaches in ECG classification. The average of sensitivity and specificity rates of detection of five arrhythmias of the second best method, \cite{ansari2016suppression}, on available dataset Physionet 2015, is a method that stands on expert knowledge and using three leads are around 89.8 $\%$ and 90.6 $\%$ while the proposed method rates are 98.42 $\%$ and 98.37 $\%$ that uses just one sensory information and completely based on machine learning. The proposed approach  in \cite{ghazanfari2019simultaneous} cannot scale up well in \cite{PhysioNet17}.
In the AFib dataset, the score as the average of F1 of detection of 3 classes of the best method on available dataset Physionet 2017 is 91.56 while the proposed method archives 98.78 in a generic form. 
%\magenta{again, this is not a fair comparison when using an augmented dataset}

\section{Conclusion}

%n general, the learning approaches with memories such as RNNs and its variations (e.g., LSTM) present efficient performance in time series analysis noting their capability in learning the temporal relations in data. However, 

The effectiveness of current approaches for ECG classification highly depends on the length of the signal, the number of available training instances, and the complexity of the patterns in a dataset. In this paper, we proposed a piece-wise matching layer as a novel approach for ECG classification that acts based on two levels. The first-level provides some local features of each periodic part and its neighbors. The second-level provides a larger view of neighbors, receptive field, to obtain higher-level features for each periodic part. The functions of the second-level are statics measures of matching between the current template of the current beat to a receptive field of other ones that provide higher-level features. In this paper, we define four scenarios including offline processing, online processing, fixed-length sliding receptive field, and event triggering receptive field.  DTW is a simple mechanism that measures the similarity of the pieces that triggers the end of the current receptive field. DTW provides abstract forms, local and dynamic perceptive processing of signal based on the events happening on a signal. We evaluated the performance of the proposed layer in two different publicly known available ECG datasets. These datasets carry known challenges to highlight the effectiveness of this layer in challenging cases where (i) the datasets are highly contaminated by noise, (ii) the datasets are imbalanced,  (iii) the length of instances varies from one to another, and (IV) there are time series with the length 70 thousand time steps. The proposed layer in all of the four scenarios considerably improves the classification performance.  Event triggering receptive achieves the best results  in dealing with noisy signals. The fixed-length sliding receptive field shows the best results in three classes of the AFib and the challenge score of the AFib. The proposed layer boosts a representation learning method as a generic framework that learns the features without any prior knowledge from human experts that can lead to considerably better performance compared to the state-of-the-art methods in both of the challenges.

\section{Appendix}
\label{sec:supplementary}
In this section, we first present a description of patients in the false alarm challenge. Then, we bring some of the formulas of the features that are extracted as morphological and statistical. Then, we brought some figures of the samples of the false alarm arrhythmia's and AFib classes.

\subsection{Description of patients in False Alarm}
In Table. \ref{other_sample}, we presented the list of the patients that the ECG lead II is not available for them, and the substitution lead which is used instead.
\begin{table}
\caption{The list of samples where the ECG lead II is not available for them \cite{ghazanfari2019simultaneous}} 

\resizebox{1\columnwidth}{!}{
\begin{tabular}{|c|c|c|}
\hline
Sample ID  & ECG Lead  & \# Patients \\
\hline\hline
b349l, b672s, b824s, &  & \\
 t116s, t208s, t209l, & I&9\\
 v289l, v290s, v619l &  & \\
\hline
t693l & aVF & 1\\
\hline
t477l, t478s, t622s, &  & \\
t665l, a457l, a582s, & V&11\\
a661l, t739l, v328s, &  & \\
v459l, v629l & &\\
\hline
a675l & III & 1 \\
\hline
\end{tabular}
}
\label{other_sample}
\end{table}

\subsection{Feature Extraction Formula}
In this section, we first present morphological formula such as Bazett, Friderica, and Sagie that are applied on QT. Bazett  \cite{bazett1997analysis}, Fridericia \cite{moss2003introductory}, and Sagie \cite{sagie1992improved} formulas consider QT interval regarding by the heart rate in different ways. 

$$
Bazett's\;QT\;formula = \frac{QT}{\sqrt{\frac{RR}{(1s)}}}
$$

$$
Friderica's\;QT\;formula= \frac{QT}{\sqrt[3]{\frac{RR}{(1s)}}}
$$

$$
Sagie's\;QT\;formula = 1000(QT/1000 + 0.154(1-RR)
$$

Then,  Hjorth parameters \cite{hjorth1970eeg} including Hjorth activity, Hjorth mobility, and Hjorth complexity that are belong to statistical group are presented.  Hjorth activity parameter shows the variance of time series as a variable that shows the surface of power spectrum in the frequency domain \cite{oh2014novel}. Hjorth mobility capture a form of standard deviation of power spectrum \cite{oh2014novel} and  Hjorth complexity shows how much the shape of a signal is similar to a sine wave \cite{oh2014novel}.

$$
Hjorth\;Activity = var(y(t))
$$

%\subsubsection{Hjorth Mobility}
$$
Hjorth Mobility = \sqrt{\frac{var(\frac{dy(t)}{dt})}{var(y(t))}}
$$

$$
Hjorth\;Complexity =  \frac{Mobility(\frac{dy(t)}{dt})}{Mobility(y(t))}
$$

Finally, three known enropies including Shannon, Tsaillis, and Renyi formula are shown in the following. In Tsaillis's entropy formula $q$ is the entropic index, which is set to 2 in this paper and in Renyi $\alpha$ is the order, which is set to 2 in this paper.

$$
Shannon\;Entropy = -\sum_{i}^{} p_i log(p_i)
$$

$$
Tsaillis\;Entropy = \frac{1}{q - 1} (1-\sum_{i}{}{p_i}^q)
$$

$$
Renyi\;Entropy  = \frac{1}{1 - \alpha} log(\sum_{i}{}{p_i}^\alpha)
$$

\subsection{False Alarm Reduction Rhythm's Samples}

In the following, we show one sample of each arrhythmia that their description explained in Table. \ref{table:2015_summary} with the patients' id of the False Alarm Reduction datasets of \cite{PhysioNet15}. 

\begin{figure}[H]
\centering
\includegraphics[width=0.9\columnwidth]{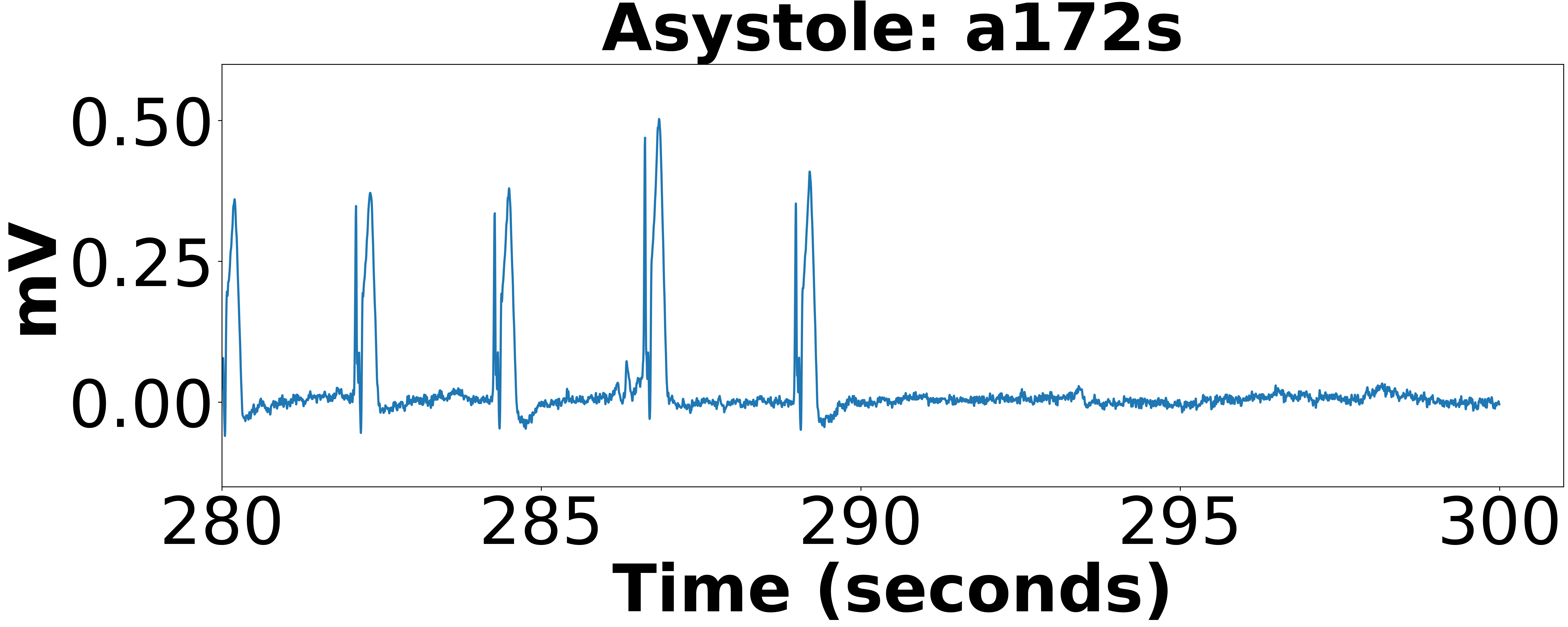}
\caption{An example of ASY rhythm. }
\end{figure}

\begin{figure}[H]
\centering
\includegraphics[width=0.9\columnwidth]{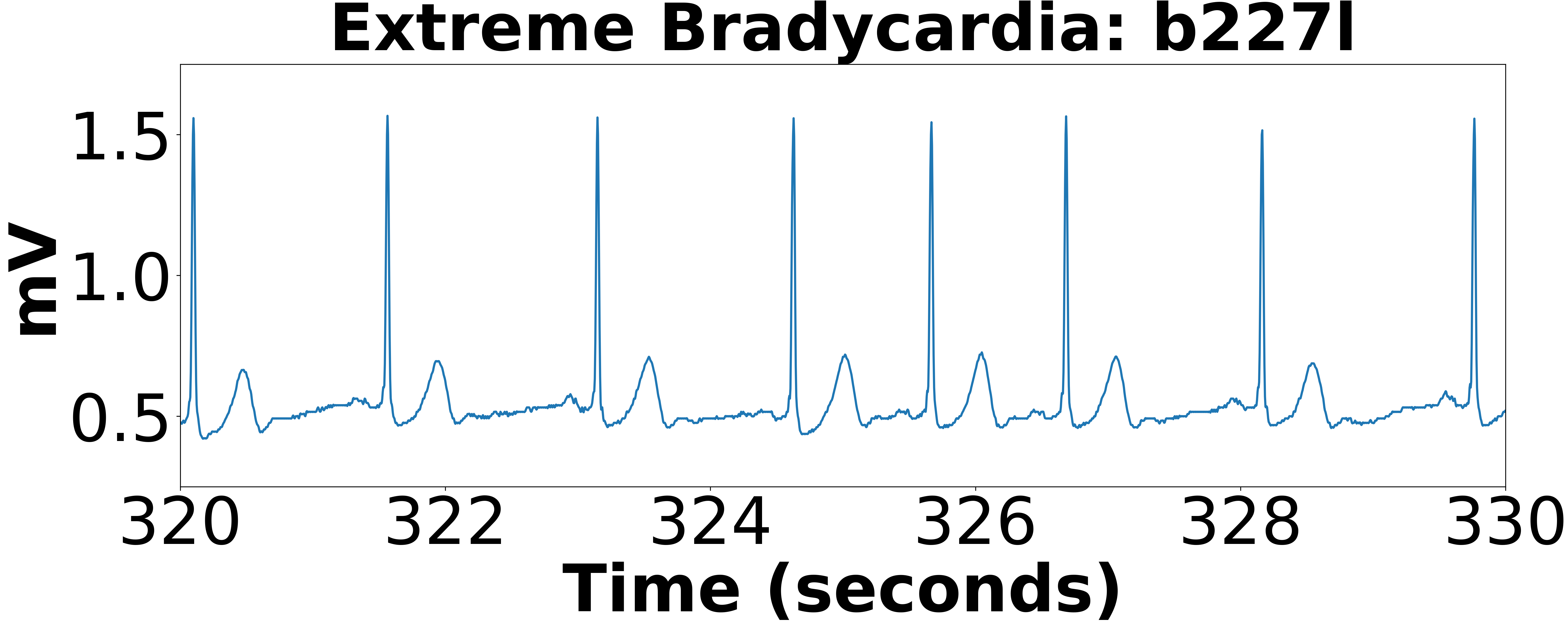}
\caption{An example of EBR rhythm.}
\end{figure}

\begin{figure}[H]
\centering
\includegraphics[width=0.9\columnwidth]{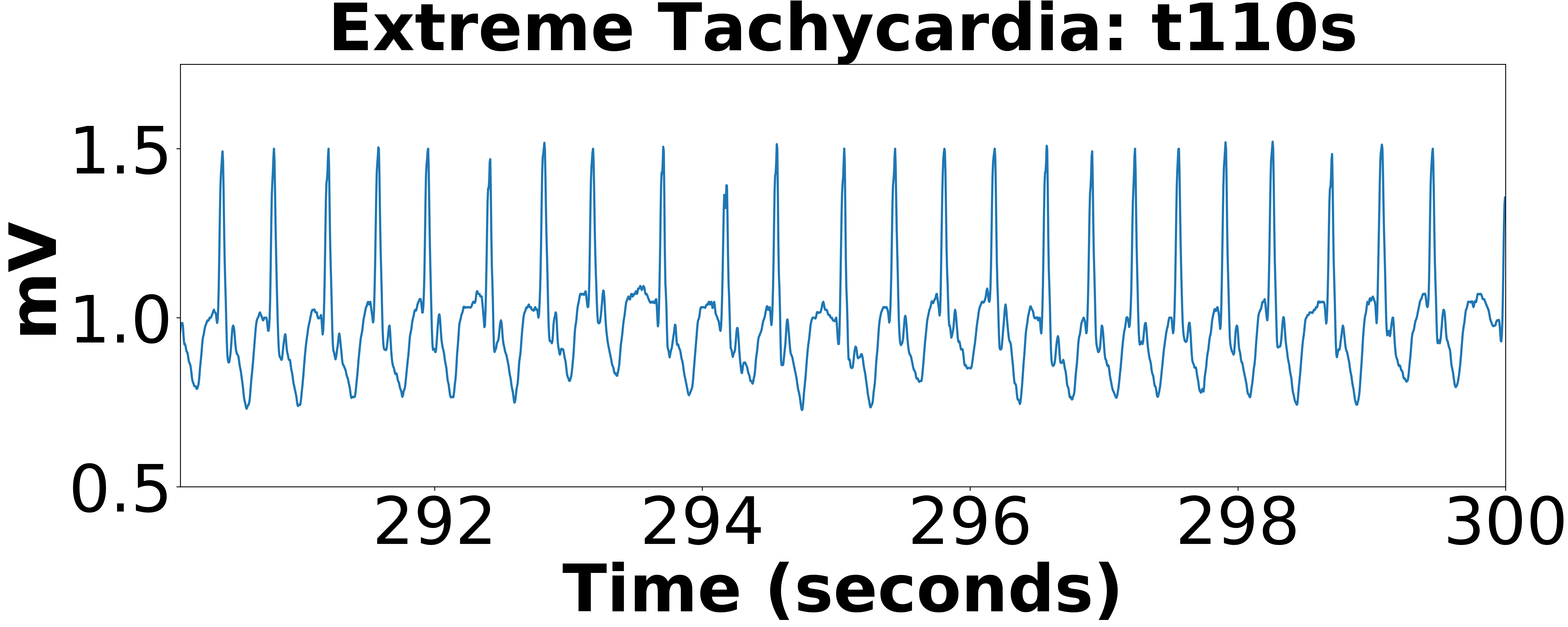}
\caption{An example of ET rhythm.}
\end{figure}

\begin{figure}[H]
\centering
\includegraphics[width=0.9\columnwidth]{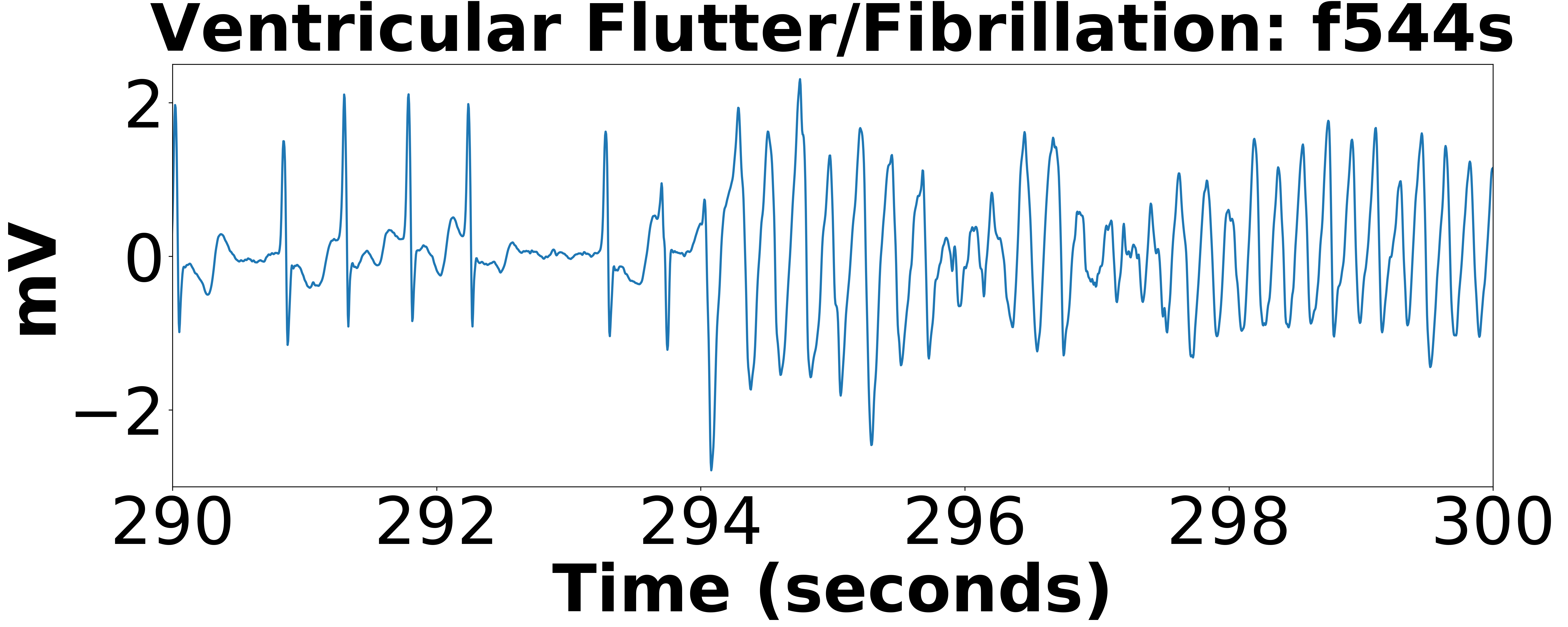}
\caption{An example of VF arrhythmia.}
\end{figure}

\begin{figure}[H]
\centering
\includegraphics[width=0.9\columnwidth]{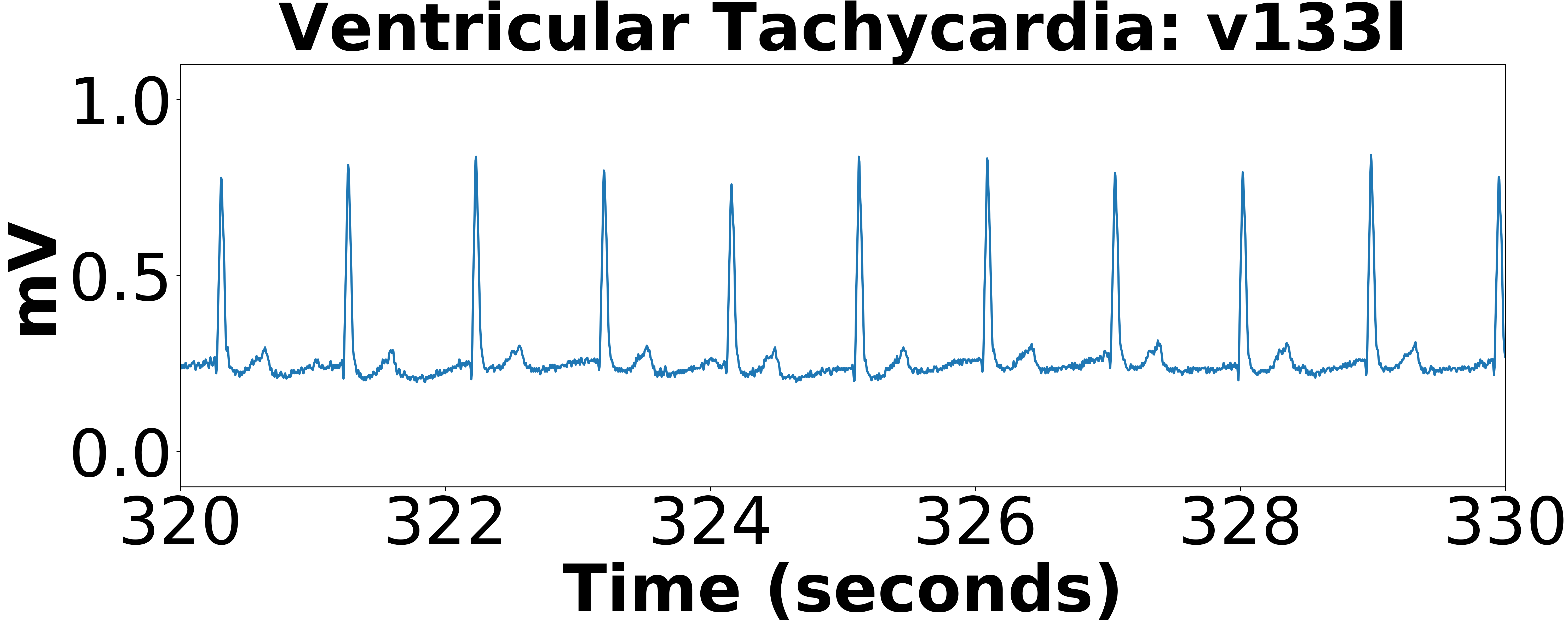}
\caption{An example of VT rhythm.}
\end{figure}

\subsection{AFib classification Samples}
In this subsection, we present some samples of each class including Normal, AFib, and Noise and some known types of the other class including \textit{Atrial flutter}, \textit{Bradycardia}, \textit{Long PR interval}, \textit{Presence of ventricular or fusion beats} and etc., in the following based on the dataset in \cite{PhysioNet17}.

\begin{figure}[H]
\centering
\includegraphics[width=0.9\columnwidth]{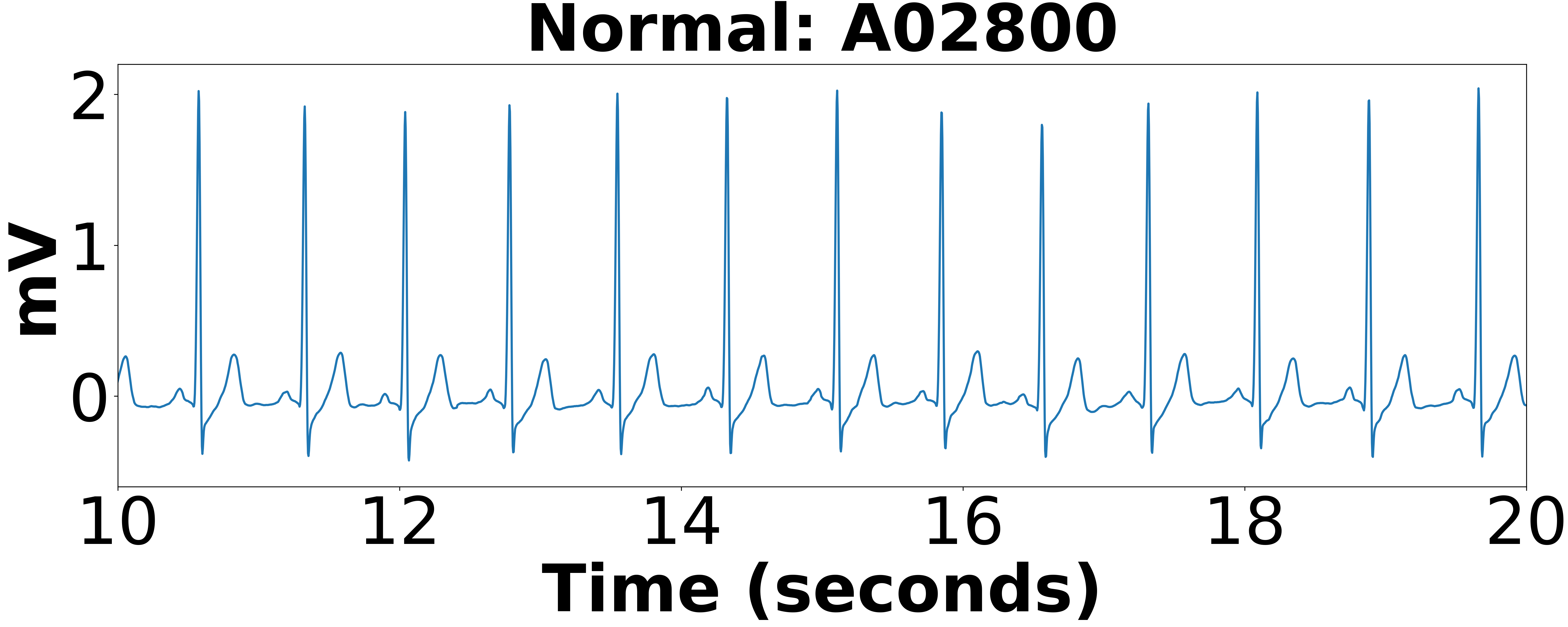}
\caption{An example of normal rhythm.}
\end{figure}

\begin{figure}[H]
\centering
\includegraphics[width=0.9\columnwidth]{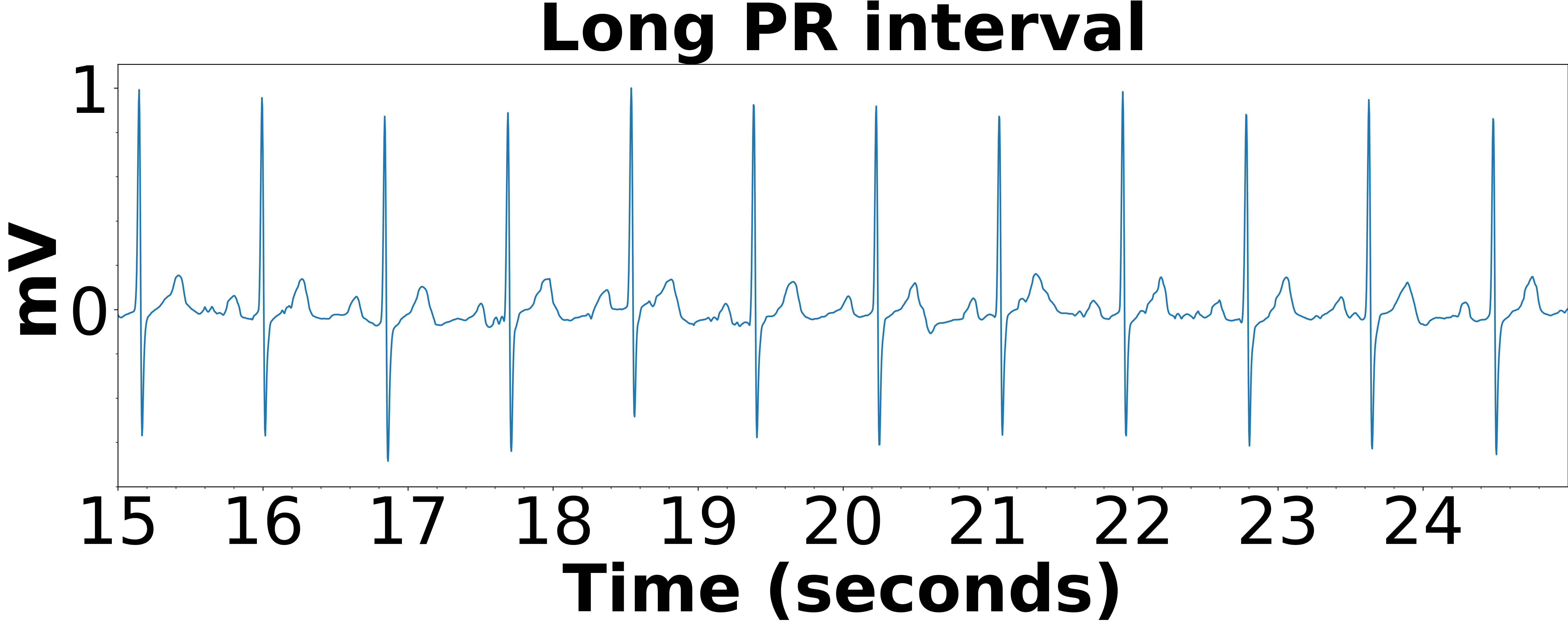}
\caption{An example of long PR interval in normal rhythm.}
\end{figure}

\begin{figure}[H]
\centering
\includegraphics[width=0.9\columnwidth]{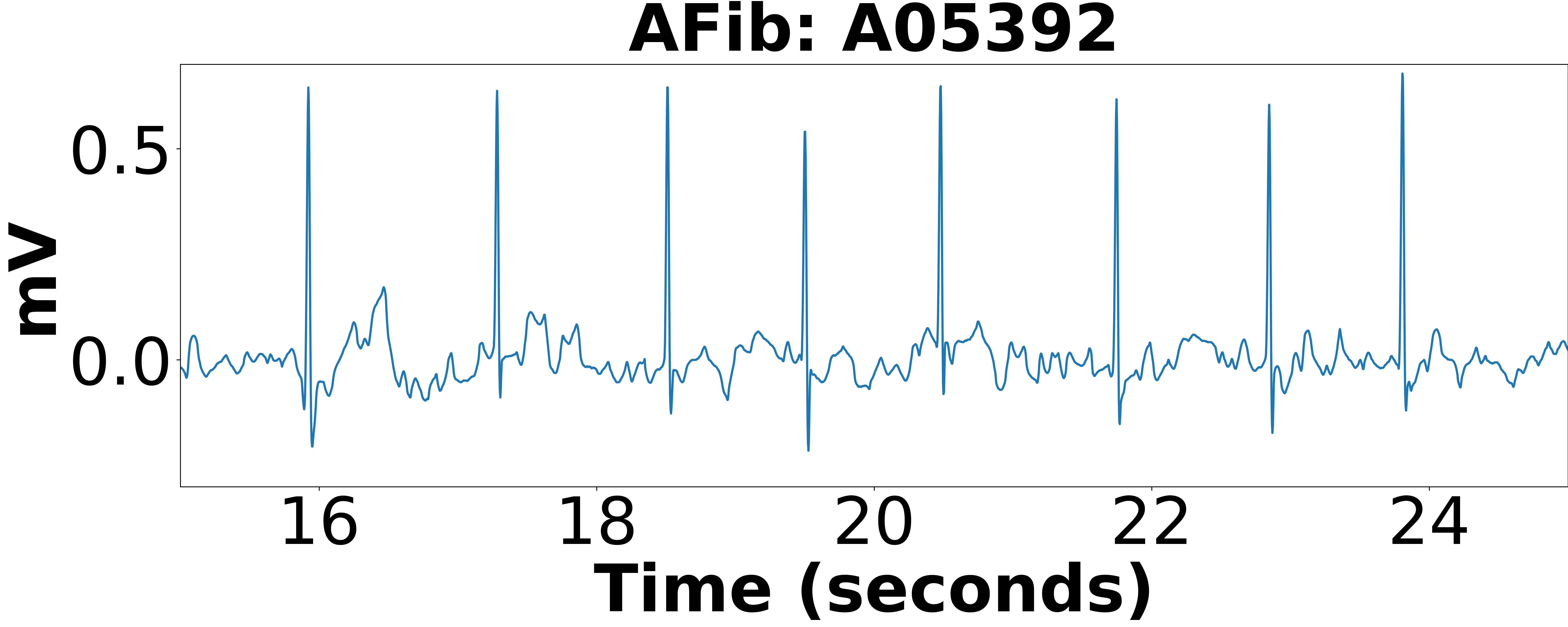}
\caption{An example of AFib rhythm.}
\end{figure}

\begin{figure}[H]
\centering
\includegraphics[width=0.9\columnwidth]{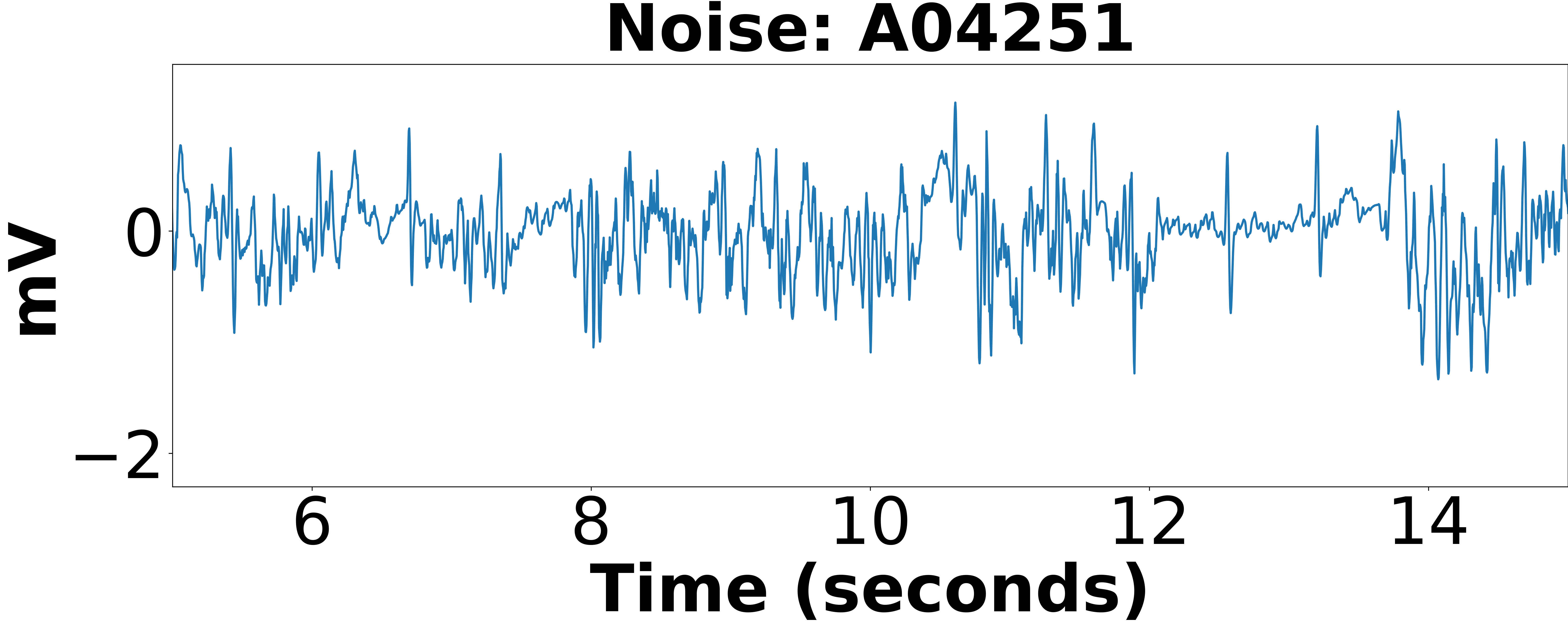}
\caption{An example of noise signal.}
\end{figure}

\begin{figure}[H]
\centering
\includegraphics[width=0.9\columnwidth]{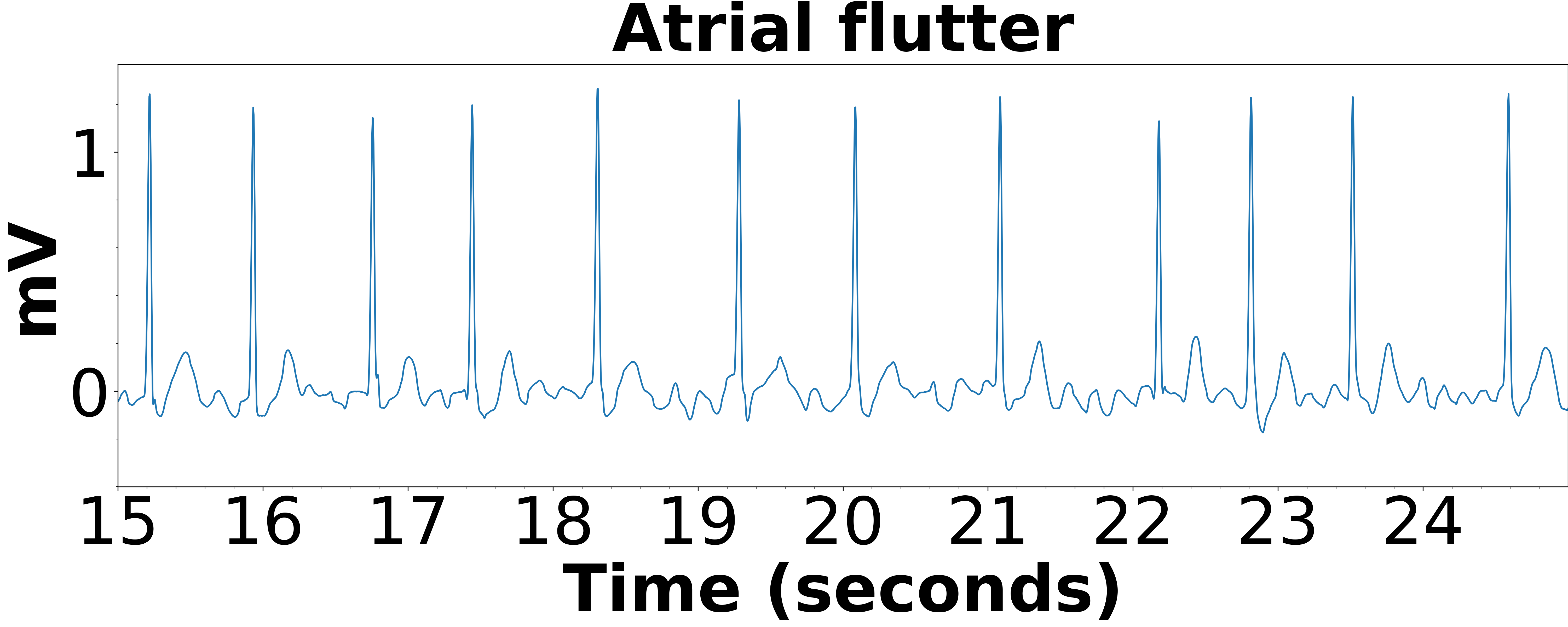}
\caption{An example of atrial flutter in other rhythm.}
\end{figure}

\begin{figure}[H]
\centering
\includegraphics[width=0.9\columnwidth]{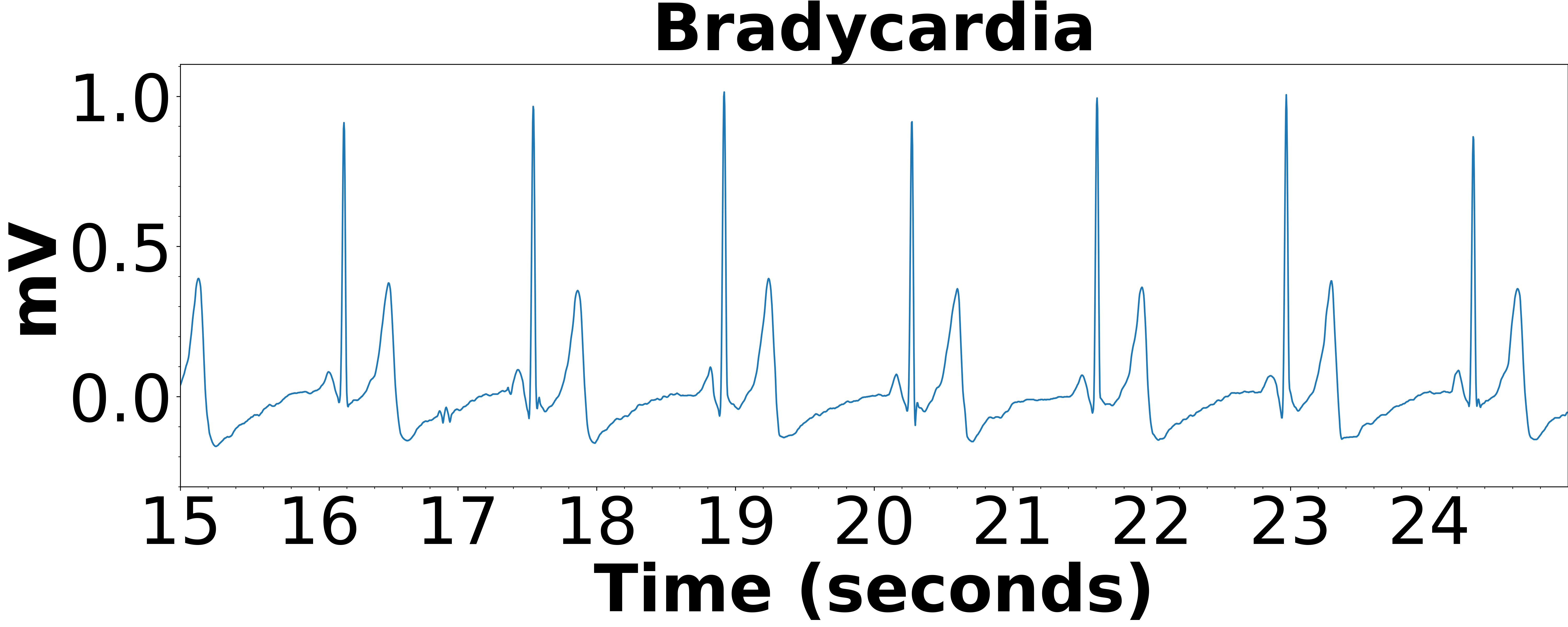}
\caption{An example of bradycardia in other rhythm.}
\end{figure}

\begin{figure}[H]
\centering
\includegraphics[width=0.9\columnwidth]{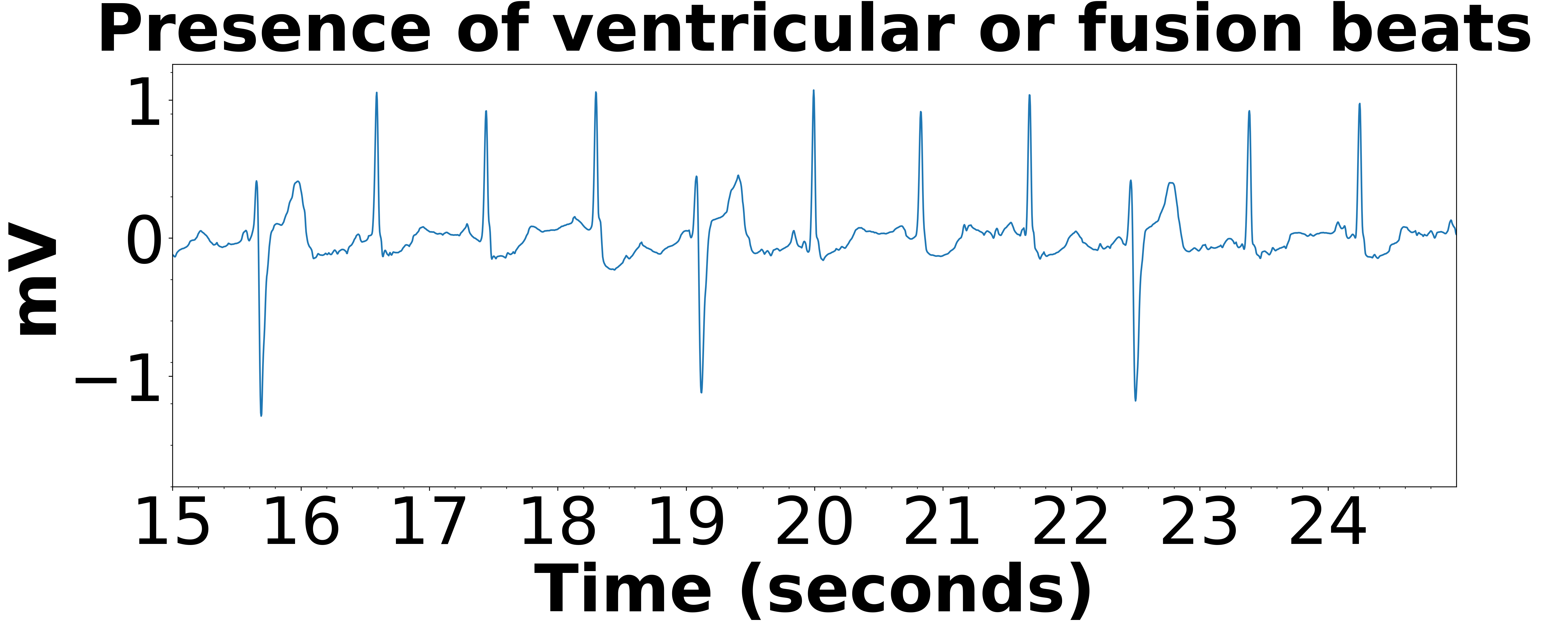}
\caption{An example of presence of ventricular or fusion beats in other rhythm.}
\end{figure}

\begin{figure}[H]
\centering
\includegraphics[width=0.9\columnwidth]{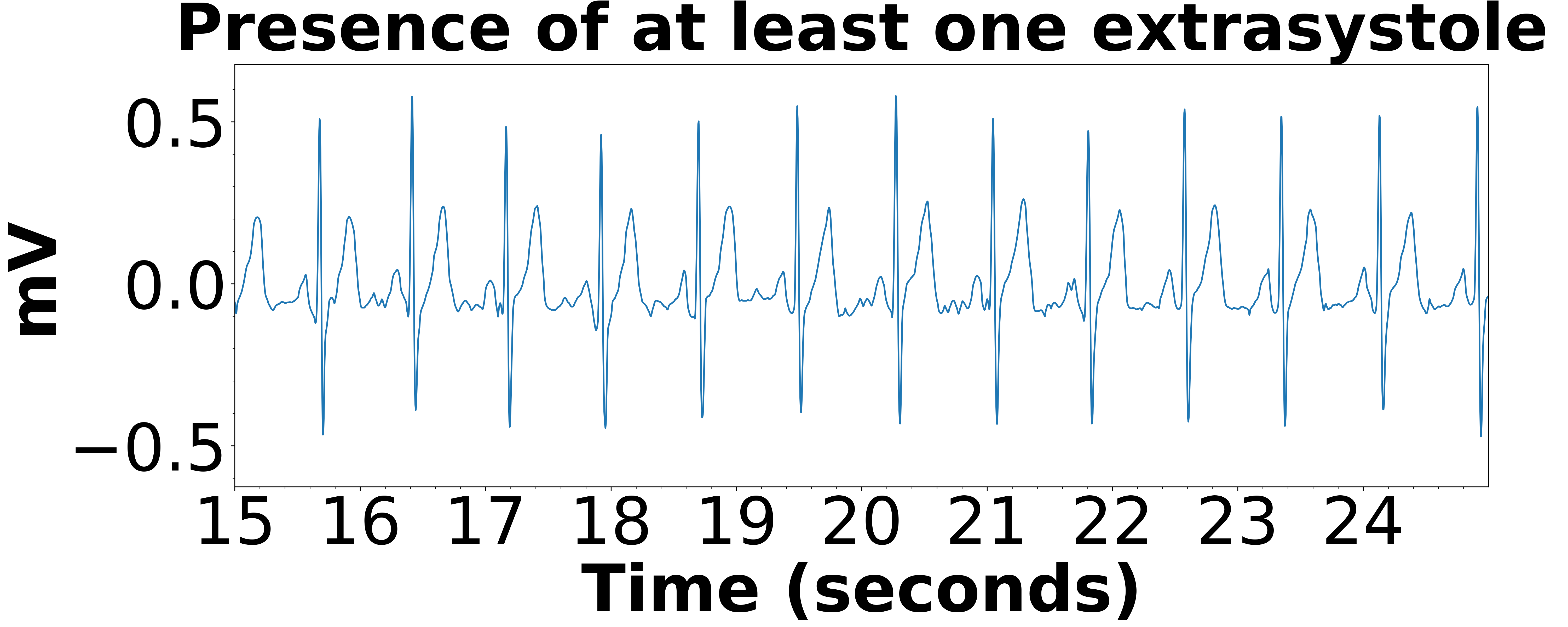}
\caption{An example of presence of at least one extrasystole.}
\end{figure}

\begin{figure}[H]
\centering
\includegraphics[width=0.9\columnwidth]{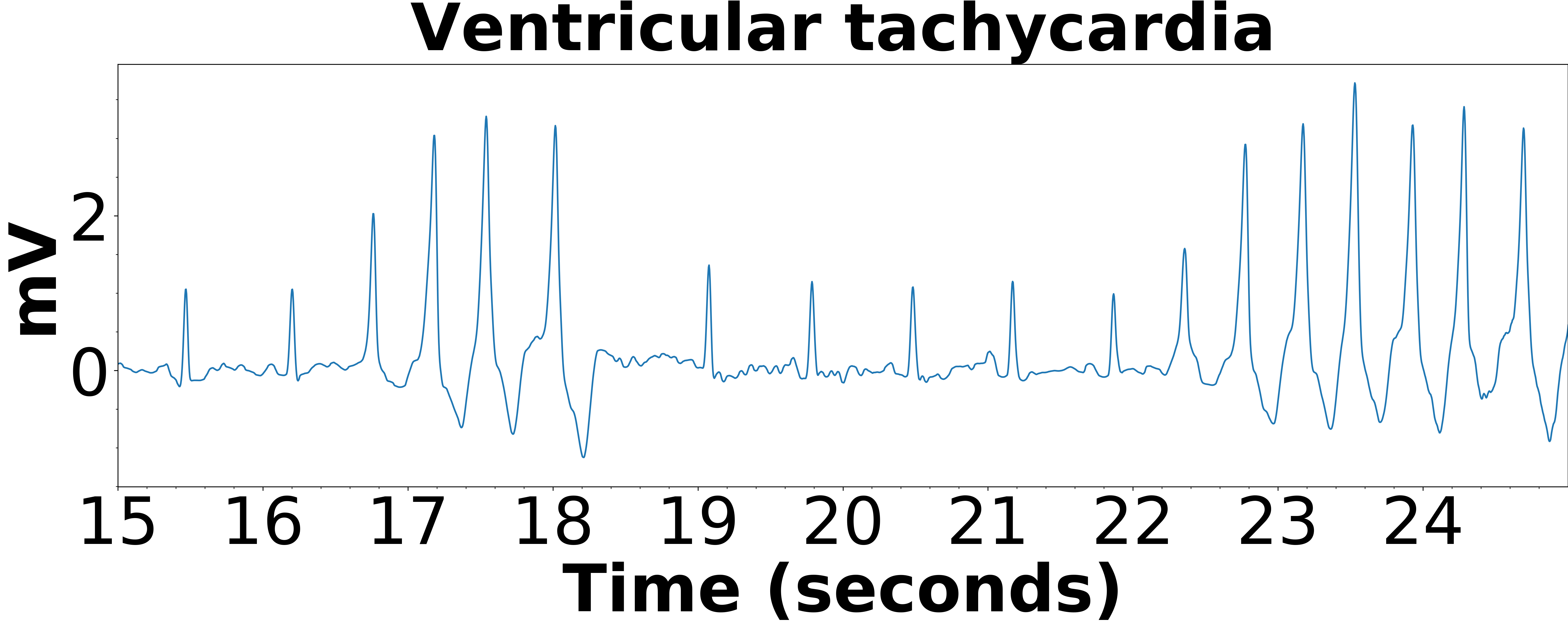}
\caption{An example of ventricular tachycardia}
\end{figure}

\begin{figure}[H]
\centering
\includegraphics[width=0.9\columnwidth]{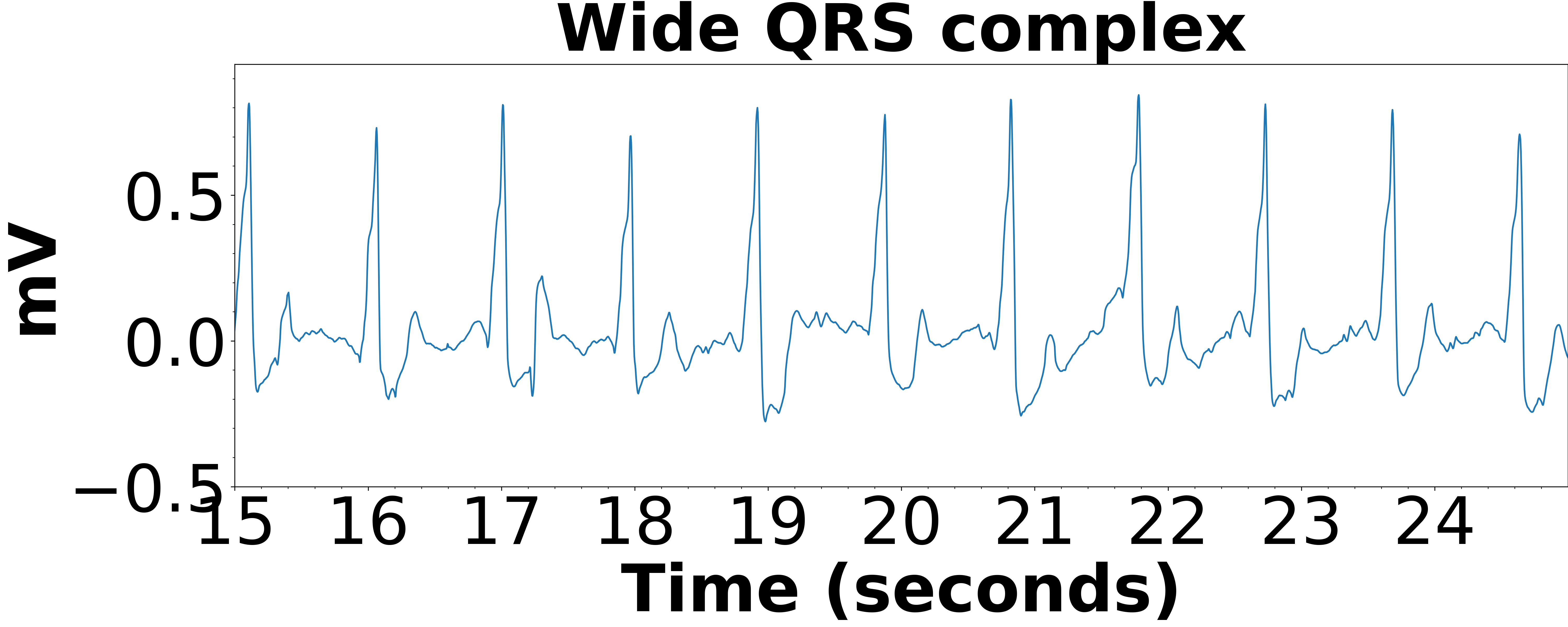}
\caption{An example of wide QRS complex.}
\end{figure}

\section*{Acknowledgment}
We gratefully thank the high performance computing
team of Northern Arizona University. We acknowledge the support of NVIDIA Corporation with the donation of the Quadro P6000 used for this research.

\bibliographystyle{IEEEtran}
\bibliography{references}

\end{document}